\documentclass[twocolumn]{aastex631}
\usepackage{makecell} 
\usepackage{comment}

\newcommand{\ntwohplus}{N$_2$H$^+$ }
\newcommand{\ntwohplusnospace}{N$_2$H$^+$}
\newcommand{\dcoplus}{DCO$^+$ }
\newcommand{\dcoplusnospace}{DCO$^+$}
\newcommand{\htwodplus}{H$_{2}$D$^{+} $ }
\newcommand{\htwodplusnospace}{H$_{2}$D$^{+}$}
\newcommand{\chtwodplus}{CH$_2$D$^{+}$ }

\graphicspath{{./}{figures/}}

\begin{document}

\title{Peering into the Midplane of the DM Tau Disk using N$_2$H$^+$ and DCO$^+$}

\author[0009-0006-0491-6210]{Amina Diop}
\affiliation{Department of Astronomy, University of Virginia, Charlottesville, VA 22904, USA}
\correspondingauthor{Amina Diop} 
\email{pmw3zz@virginia.edu}

\author[0000-0003-2657-1314]{Kevin Flaherty}
\affiliation{Department of Astronomy and Department of Physics, Williams College, Williamstown, MA,01267,USA}

\author[0000-0001-8642-1786]{Chunhua Qi}
\affiliation{Institute for Astrophysical Research, Boston University, 725 Commonwealth Avenue, Boston, MA 02215, USA}

\author[0000-0003-2076-8001]{Ilse Cleeves}
\affiliation{Department of Astronomy, University of Virginia, Charlottesville, VA 22904, USA}

\author[0000-0002-4803-6200]{A. Meredith Hughes}
\affiliation{Department of Astronomy, Van Vleck Observatory, Wesleyan University, 96 Foss Hill Dr., Middletown, CT 06459, USA}

\author[0000-0001-7157-6275]{\'Agnes K\'osp\'al}
\affiliation{Konkoly Observatory, HUN-REN Research Centre for Astronomy and Earth Sciences, MTA Centre of Excellence, Konkoly-Thege Mikl\'os \'ut 15-17, 1121 Budapest, Hungary}
\affiliation{Institute of Physics and Astronomy, ELTE E\"otv\"os Lor\'and University, P\'azm\'any P\'eter s\'et\'any 1/A, 1117 Budapest, Hungary}
\affiliation{Max-Planck-Insitut f\"ur Astronomie, K\"onigstuhl 17, 69117 Heidelberg, Germany}

\author[0000-0003-2253-2270]{Sean Andrews}
\affiliation{Center for Astrophysics, Harvard \& Smithsonian, 60 Garden Street, Cambridge, MA 02138-1516, USA}

\author[0000-0003-1526-7587]{David J. Wilner}
\affiliation{Center for Astrophysics, Harvard \& Smithsonian, 60 Garden Street, Cambridge, MA 02138-1516, USA}

\author[0000-0002-3771-8054]{Jacob B. Simon}
\affiliation{Department of Physics and Astronomy, Iowa State University, Ames, IA 50010, USA}

\author[0000-0003-3840-7490]{Deryl E. Long}
\affiliation{Department of Astronomy, University of Virginia, Charlottesville, VA 22904, USA}

\shorttitle{Diop et al.}
\shortauthors{Diop et al.}

\begin{abstract} 
Planets form in the midplanes of protoplanetary disks, thus investigating the properties of the midplane region can shed light on the physical and chemical processes governing the evolution of planetary systems.  In this work, we study the conditions near the midplane of the disk around DM Tau using 0$\farcs$3 ($\sim$ 43 au) ALMA observations   of  \ntwohplus ($J=3-2$) and \dcoplus($J=4-3$).
The radial profiles reveal that \dcoplus is centrally peaked, with a depression near 120 au, while \ntwohplus exhibits a central depression and a peak near 100 au. In addition, both molecules display extended emission out to 400 au. Using parametric models to interpret our observations, we note that a fiducial model based on CO-derived parameters captures the features within 200 au, but fails to reproduce the outer disk emission. We explore various mechanisms that could account for this extended emission and conclude that it is best reproduced by an increased external UV penetration and cosmic ray/X-ray ionization rate, possibly enhanced by radial drift. These processes affect the precursors of \ntwohplus and \dcoplus: increased UV leads to the photodesorption of CO and N$_2$, while ionization enhances H$_3^+$ abundance. Besides, the observed \ntwohplus and \dcoplus features are broadly consistent with previous CO snowline constraints, but can also be shaped by processes such as ionization, CO depletion, and multiple chemical pathways. Thus, using these molecular ions as robust CO snowline tracers requires additional constraints from complementary molecular tracers and detailed chemical modeling. 
\end{abstract}
\keywords{Protoplanetary disks --- Stars: individual (DM Tau) --- Astrochemistry} 

\section{Introduction} \label{sec:intro}
Protoplanetary disks constitute the birthplace of planetary systems. Their chemical makeup dictates the composition of nascent planets and offers insights into planet formation, evolution, and migration histories  \citep[e.g.,][]{oberg11,madhu19,oberg2023,zhang24}. Molecular line emission has been used to constrain various disk properties such as the gas density \citep[e.g.,][]{zhang2021,schwarz2021}, temperature \citep[e.g.,][]{kama2016,pinte2018,calahan2021}, ionization \citep[e.g.,][]{cleeves2015,seifert2021,aikawa2021,Long24},  and turbulence \citep[e.g.,][]{Flaherty2017, Flaherty2020,paneque24}. 

Among the molecules observed in disks, \ntwohplus and \dcoplus have been frequently used to probe the environment near the dense midplane region, where planets form \citep[e.g.,][]{willacy07,walsh10,qi19,munozromero}. These molecular ions exhibit distinct radial emission morphologies that can be analyzed to infer the underlying physical and chemical conditions. In the case of \ntwohplusnospace, it is often concentrated in a ring \citep[e.g.,][]{qi132,qi13,qi19}, a morphology attributed to its formation from N$_2$ gas and destruction by CO gas \citep{bergin01,jorgensen04,van2017}. In other words, \ntwohplus comes from the regions where CO is frozen out and N$_2$ is still in the gas phase. Given the dependence on CO and N$_2$ freeze-out temperatures, the thermal structure of the disk is a key factor regulating the spatial distribution of \ntwohplus \citep{van2017}. \citet{qi19} find that disks can be characterized by two distinct \ntwohplus morphologies depending on their temperature structure.  Some disks (DM Tau, GM Aur, LkCa15) show a bright, narrow ring produced by a thick Vertically Isothermal Region above the Midplane (VIRaM) that concentrates \ntwohplus between the midplane CO and N$_2$ snowlines, accompanied by tenuous extended emission arising from the low-density upper layers in the outer disk. In contrast, other disks (V4046 Sgr, AS 209, IM Lup) display a radially broad emission due to a thin VIRaM which confines \ntwohplus to vertically thin layers distributed over a range of radii. They also report secondary rings in LkCa 15, GM Aur, IM Lup close to the mm dust edge that could be indicative of increased penetration of radiation, leading to a temperature enhancement and/or increased ionization.

As for \dcoplusnospace, it has been observed in several disks and is usually characterized by a series of rings \citep[e.g.,][]{oberg2015,teague15, Flaherty2017,huang17,salinas17,carney18,kastner18,obergrosetta,munozromero}. This multi-ring structure may be due to the various formation pathways of \dcoplus (for more details, see Section \ref{sec:chemistry}): for example, \dcoplus can be formed via a warm pathway and a cold pathway. The former dominates at $30 \; \text{K}<\text{T} < 70  \; \text{K}$ and involves \chtwodplus or HCO$^+$ as reactants \citep{wooten87,millar1989, favre2015}, while the latter dominates at T $<$ 30 K with its reactants being \htwodplus and cold CO gas \citep{millar1989}. Interestingly, the outermost rings of \dcoplus are typically observed in the cold outer disk where CO is expected to be frozen out closer to the midplane, which would inhibit the formation of \dcoplus  via the cold pathway. Their presence suggests that a fraction of CO is returned to the gas phase, potentially through non-thermal desorption driven by UV photons penetrating the low-density regions in the outer disk and/or thermal desorption enhanced by radial drift \citep[e.g.,][]{oberg2015,Cleeves2016a}. These results demonstrate that \ntwohplus and \dcoplus are sensitive to the disk thermal profile, gas-phase CO abundance, dust evolution, and the degree of radiation penetration, making them valuable probes of the physical and chemical processes operating close to the disk midplane where planets form. 

In this work, we study the region near the midplane of the DM Tau disk using \ntwohplus and \dcoplusnospace. DM Tau is a 5-Myr old T Tauri star \citep{Simon2000} located $\sim$ 144.5 pc away \citep{Lindegren2018,bailer18} in the Taurus star-forming region and is surrounded by a transition disk with dust rings at $\sim$ 4 and $\sim$ 19 au \citep{bergin04,calvet05,Andrews2011,Kudo2018}. The DM Tau disk is an ideal target for probing midplane processes since some of its key properties have been previously constrained, including its CO snowline location, degree of CO depletion, temperature structure, and ionization profile  \citep{qi19,zhang2019,Flaherty2020,law2022,law2023,galloway25,Long24}. These constraints provide valuable context for assessing which processes are able to reproduce the observed  \ntwohplus and \dcoplus morphologies. By focusing on a well-characterized disk, this study provides insight into the processes that may drive N$_2$H$^+$ and DCO$^+$ emission in protoplanetary disks. This paper is organized as follows. Section \ref{sec:obs} describes our observations and data reduction procedure. In Section \ref{sec:obs_res}, we present the observational results, followed by our modeling approach in Section \ref{sec:models} and our findings in Section \ref{sec:mod_results}. Finally, we discuss the implications of our work in Section \ref{sec:disc}.

\section{Observations and Data Reduction} \label{sec:obs}
We worked with  0$\farcs$3 ($\sim$ 43 au at 144.5 pc)  ALMA continuum, N$_2$H$^+$($J=3-2$), and DCO$^+$($J=4-3$) data from the project 2018.1.01119.S (PI: K. Flaherty). The observations were carried out on October 25, November 4, and November 5 2018, with a total integration time of 235 min on the source and with baselines ranging from 15 to 1360 m. All observations used the same phase center (RA = 04:33:48.734, DEC: 18:10:09.644).  The spectral windows were centered on the rest frequencies of \ntwohplus and \dcoplus (279.511 GHz and 288.143 GHz respectively), and on 278/290 GHz for the continuum. The channel width was 244.141 kHz  for the DCO$^+$ and N$_2$H$^+$ spectral windows ($\sim $ 254 m/s for DCO$^+$ and $\sim$ 262 m/s for N$_2$H$^+$) and 15,625 kHz for the continuum.  Our observations are summarized in Table \ref{tab:obs}. 
\begin{figure*}[t!]
\centering
    \includegraphics[width=\textwidth]{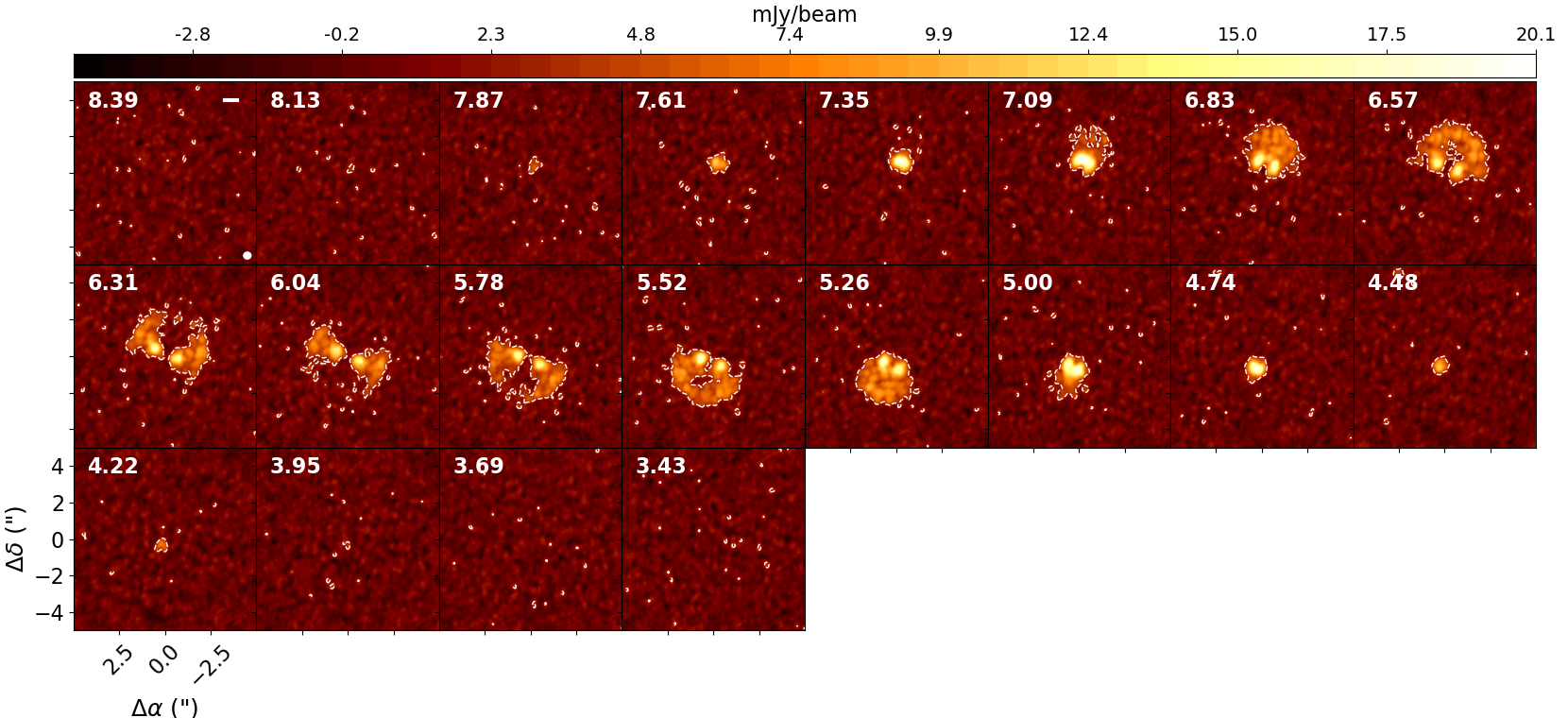}
    \includegraphics[width=\textwidth]{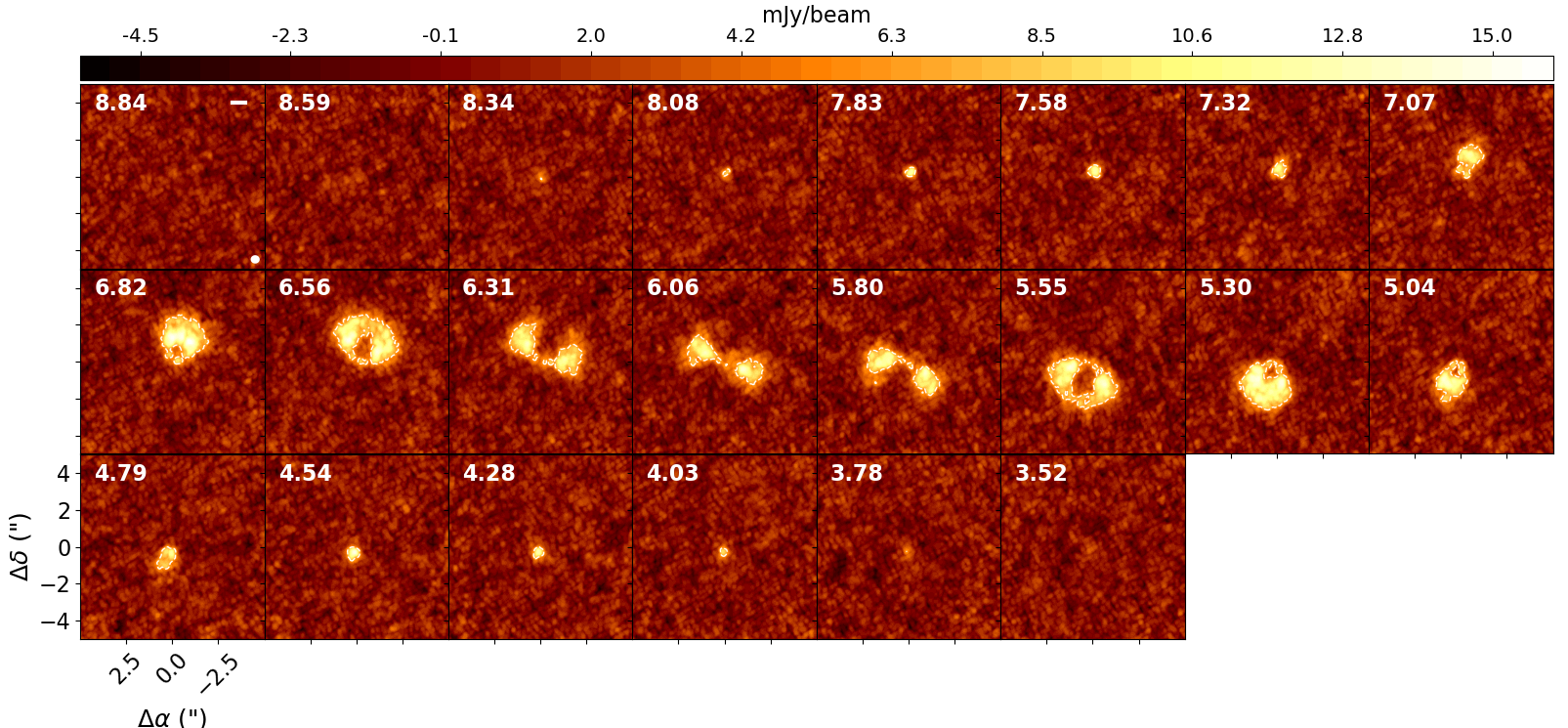}
    \caption{Full channel maps for \ntwohplus ($J=3-2$) (\textit{top}) and \dcoplus ($J=4-3$) (\textit{bottom}). Numbers in the upper left corners are the velocity of each channel in km/s. A 100 au scale bar and the beam size are shown in the upper left panel, with 3$\sigma$ contours as white-dashed lines in each panel. \label{fig:n2h_dco_all_channels}}
\end{figure*}

\begin{figure*}[hbt!]
    \plottwo{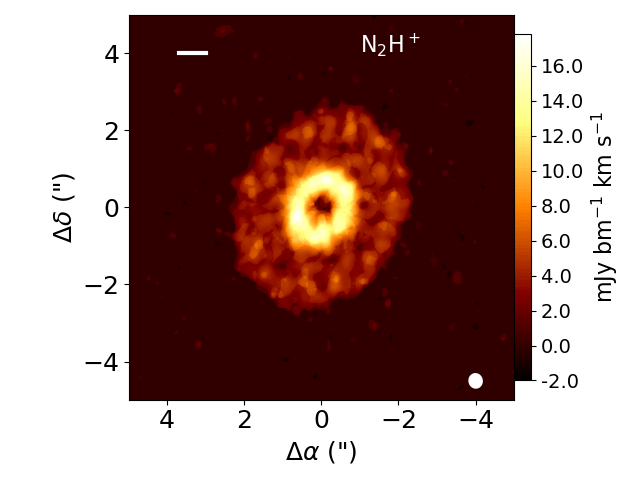}{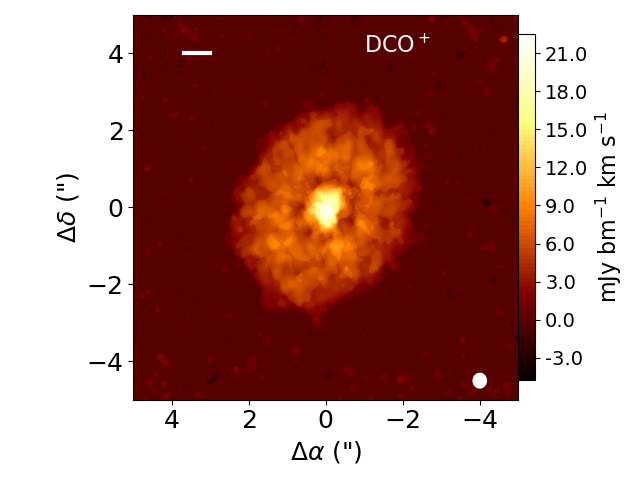}
    \caption {\textit{Left:} N$_2$H$^+$ ($J=3-2$) moment 0 map generated using channel emission above 3$\sigma$ ($\sigma$ = 1.54 mJy/beam).  The \ntwohplus emission displays an inner hole surrounded by a bright ring and tenuous extended emission. \textit{Right:} DCO$^+$ ($J=4-3$) moment 0 map  generated using channel emission above 3$\sigma$ ($\sigma$  = 1.24 mJy/beam). The \dcoplus emission is characterized by a bright central region surrounded by extended emission. In both panels, the white line and circle are respectively a 100 au scale bar and the synthesized beam.}
    \label{fig:n2h_dco_mom_maps}
\end{figure*}

For the data reduction, we used the Common Astronomy Software Applications package known as CASA \citep{mcmullin07}. The calibration and flagging were performed using the standard ALMA pipeline. One round of phase-only self-calibration, using a solution interval of 30s, was calculated and applied to each execution block independently; additional rounds of self-calibration did not significantly improve the SNR. We then used the \texttt{mstransform} task to combine the data from different epochs into a common velocity scale and the \texttt{uvcontsub} task to remove the continuum from the line emission. We selected channels with significant line emission and used the \texttt{tclean} task with Briggs weighting and a robust parameter of 1.5 to create the channel maps (Figure \ref{fig:n2h_dco_all_channels}). 
For N$_{2}$H$^{+}$, the beam size and the rms level are respectively  0.$\arcsec$35 $\times$ 0.$\arcsec$32 and 1.54 mJy/beam, while for DCO$^{+}$, they are 0.$\arcsec$34 $\times$ 0.$\arcsec$31 and 1.24 mJy/beam. We used the \texttt{bettermoments} package \citep{Bettermoments2018,Teague2019} to generate moment 0 maps from the channel maps (Figure \ref{fig:n2h_dco_mom_maps}). 
\vspace{-7mm}

\begin{deluxetable*}{ccccccc}[t!]
\tablecaption{Observing Summary\label{tab:obs}}
\tablewidth{\textwidth}
\tablehead{{Dates} & \colhead{Number of Antennas} & \colhead{Integration Time} & \colhead{Baseline lengths} & \colhead{Calibrators} & \colhead{PWV (mm)}}
\decimals
\startdata
\vspace{0.3cm}
\makecell{2018-10-25 \\ 05:12 - 06:28} & 47 & 47 min & 15 - 1360 m & \makecell{J0423-0120\\ J0501+1356\\ J0510+1800} & 0.6 \\
\vspace{0.3cm}
\makecell{2018-11-04 \\ 07:09 - 08:25} & 46 & 47 min & 15 - 1360 m & \makecell{J0423-0120\\ J0510+1800 \\J0530+1331} & 1.6\\
\vspace{0.3cm}
\makecell{2018-11-04 \\ 08:27 - 09:43} & 46 & 47 min & 15 - 1360 m & \makecell{J0440+1437\\ J0449+1121\\ J0510+1800}  & 1.4 - 1.7 \\
\vspace{0.3cm}
\makecell{2018-11-05 \\ 04:14 - 05:30} & 46 & 47 min & 15 - 1360 m & \makecell{J0431+1731\\ J0440+1437\\ J0510+1800} & 1.0 \\
\vspace{0.3cm}
\makecell{2018-11-05 \\ 05:33 - 06:49} & 46 & 47 min & 15 - 1360 m & \makecell{J0431+1731 \\J0440+1437\\ J0510+1800} & 0.8 \\
\enddata
\end{deluxetable*}

\section{Observational results}
\begin{figure}[htb!]
      \begin{center}
      \includegraphics[width=0.43\textwidth]{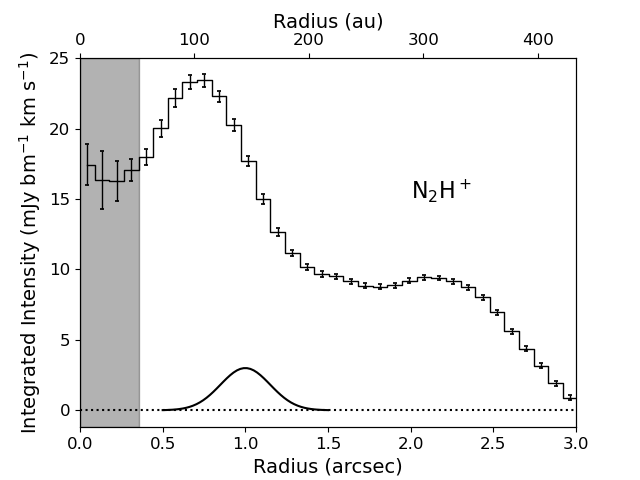}
      \includegraphics[width=0.43\textwidth]{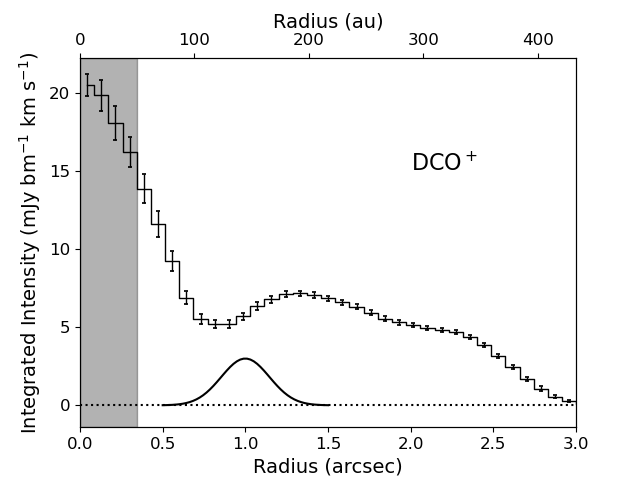}
      \includegraphics[width=0.43\textwidth]{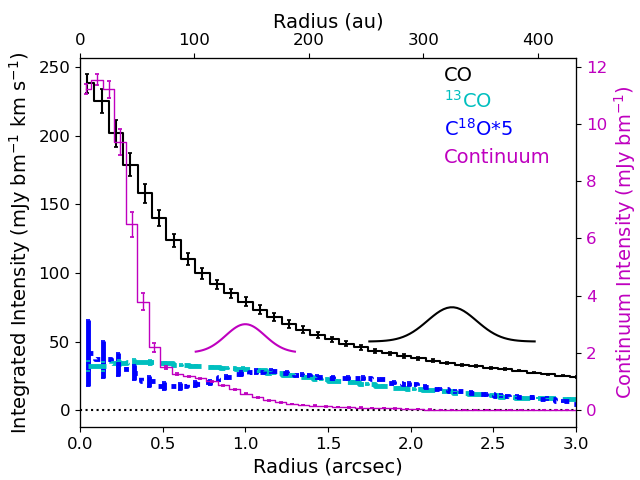}
      \end{center}
\caption {\textit{Top and middle:} Radial profiles for N$_2$H$^+$ and DCO$^+$. The solid Gaussian shows the beam shape and the grey band indicates the region that is the most sensitive to our choice of sigma clipping. \dcoplus peaks at the center and \ntwohplus near 100 au.  \textit{Bottom:} Radial profiles for CO ($J=2-1$) (thick black line), $^{13}$CO (J=2--1) (cyan dashed line), C$^{18}$O (J=2--1) (blue dotted line, scaled up by a factor of five to increase visibility) and the mm continuum emission (thin magenta line) \citep{Flaherty2020}. CO/$^{13}$CO/C$^{18}$O}, N$_2$H$^+$, and DCO$^+$ emission extends out to $\sim$ 400 au, while the continuum is compact ($\sim$ 200 au).
\label{fig:n2h_dco_co_cont_rad_profs}
\end{figure}

{\label{sec:obs_res}}
Figure \ref{fig:n2h_dco_co_cont_rad_profs} shows the radial profiles for N$_2$H$^+$ and DCO$^+$ as derived from azimuthally averaging emission greater than 3$\sigma$ using \texttt{gofish} \citep{GoFish2019}. N$_2$H$^+$ shows a central flattening surrounded by a bright ring with a FWHM of 0$\farcs$8 and a peak at 100 au (0$\farcs$75). Beyond this bright ring of emission, there is a shelf of emission that extends out to 400 au (2$\farcs$75). The shelf may be a second broad ring with a peak at $\sim$310 au (2$\farcs$1). DCO$^+$ shows a central peak with a depression at $\sim$ 120 au (0$\farcs$82), followed by extended emission out to 420 au (2$\farcs$93). The extended emission itself shows a peak at $\sim$ 190 au (1$\farcs$33). The central emission is highly sensitive to our choice of sigma clipping; here we exclude flux $<$ 3$\sigma$, but increasing this clipping to 5$\sigma$ reduces the central emission from both N$_2$H$^+$ and DCO$^+$, making both appear to have central depressions. The region most affected by the clipping threshold is marked in grey in Figure \ref{fig:n2h_dco_co_cont_rad_profs}. We estimate the innermost emitting radius for \ntwohplus and \dcoplus using their velocity ranges from Figure \ref{fig:n2h_dco_all_channels} (3.95-8.13 km/s for \ntwohplus and 3.78-8.34 km/s for \dcoplusnospace), the stellar mass and inclination (Table \ref{tab:fixed_params}) as:
\begin{equation}
    r_{in} = \frac{GM_*}{\Delta_v^2} \; \textrm{sin}^2(\textrm{i})
\end{equation}
where $\Delta_v$ is the maximum velocity offset from the systemic velocity. This yields $r_{in}$ $\sim$ 37 au for \ntwohplus and $r_{in}$ $\sim$ 31 au for \dcoplusnospace. 

The N$_2$H$^+$ morphology is similar to that found in \citet{qi19}, although the higher signal-to-noise ratio of our data reveals a secondary peak at $\sim$310 au. Broadly speaking, the N$_2$H$^+$ morphology around DM Tau follows the bright narrow ring and extended tenuous emission structure that \citet{qi19} also saw around LkCa 15 and GM Aur. They interpret the DM Tau morphology as indicative of a thick VIRaM possibly associated with strong stirring of dust grains into the disk atmosphere, which is consistent with the non-zero turbulence in the system \citep{Flaherty2020,hardiman2026}. The central peak and extended plateau morphology of DCO$^+$ is similar to that around TW Hya, IM Lup, HD 163296, and HD 169142 \citep{Qi08,oberg12, oberg2015,huang17,Flaherty2017, carney18,munozromero}.

\section{Disk model} \label{sec:models}
\subsection{Model Description}
To uncover the processes leading to the observed features, we employed the ray-tracing radiative transfer code with a parametric model of the disk structure from \citet{Flaherty2020}, based on earlier work from \citet{Dartois2003} and \citet{Rose2013}. The gas surface density is expressed as a power law with an exponential tail as expected for a viscously evolving disk
\begin{equation}
    \Sigma_{\text{gas}}=\frac{M_{\text{gas}}(2-\gamma)}{2\pi R_c^2}\left (\frac{r}{R_c}\right )^{-\gamma} \;\text{exp} \left [- \left (\frac{r}{R_c} \right )^{2-\gamma} \right]
\end{equation}
 \citep{Bell1974, Hartmann1998} where $M_{\text{gas}}$ is the disk gas mass, $r$ the radius, $R_c$ the critical radius, and $\gamma$ the surface density index. The temperature structure is defined as a power law with radius, with a vertical gradient connecting the warm atmosphere and the cold midplane. This structure is expressed as:
\begin{equation}
        T_{\text{atm}} = T_{\text{atm0}} \left (\frac{r}{150 \;\text{au}} \right) ^{q}
\end{equation}
\begin{equation}
        T_{\text{mid}} = T_{\text{mid0}} \left (\frac{r}{150 \;\text{au}} \right)^{q}
\end{equation}

\begin{equation}
    T_{\text{gas}}(\text{r},\text{z}) = 
    \left\{
     \begin{array}{ll}
       T_{\text{atm}}+ (T_{\text{mid}} - T_{\text{atm}}) \left (\text{cos}\frac{\pi \text{z}}{2Z_q} \right)^2  &  \text{if} \; \text{z}< Z_q\\
         T_{\text{atm}} & \text{if}\; \text{z}\geq Z_q\\
    \end{array}
    \right.
\end{equation}

 \begin{equation}
       Z_{q} = Z_{q0} \left (\frac{r}{150 \text{AU}} \right)^{1.3}
\end{equation}
where $T_{\text{atm}}$ is the temperature of the atmosphere, $T_{\text{mid}}$ that of the midplane, $q$ ($<$ 0) the temperature power-law index, and $\text{Z}_q$ the height above the midplane at which the gas temperature reaches its maximum value. The spectral line profile is a Gaussian whose width depends on the thermal and non-thermal motions (Eq. \ref{spec_line}), with the latter associated with turbulence via the dimensionless  parameter $\delta_{\text{vturb}}$, defined as the ratio of the turbulent velocity dispersion to the thermal velocity of the emitting molecule,  $\delta_{\text{vturb}}  = \frac{v_{turb}}{v_{th}}$, where $v_{th} = \sqrt{(2\text{k}_{\text{B}} \text{T}_{\text{gas}})/({\text{m}_{\text{molecule}}})}$. 
\begin{equation}\label{spec_line}
    \Delta \text{V} = \sqrt{\frac{2\text{k}_{\text{B}} \text{T}_{\text{gas}}(\text{r},\text{z})}{\text{m}_{\text{molecule}}} (1+\delta_{\text{vturb}}^2)}
\end{equation}
 The gas density is derived based on the surface density, assuming hydrostatic equilibrium given the temperature profile. The velocity profile is Keplerian, with corrections for the height above the midplane and the gas pressure gradient.

\subsection{Modeling N$_2$H$^+$ and DCO$^+$}\label{sec:chemistry}
On top of the underlying density and temperature structure, we add N$_2$H$^+$ and DCO$^+$ abundance distributions informed by their chemistry. For \ntwohplusnospace, the formation and destruction pathways are
\begin{equation}\label{eq:n2h_formation}
     \text{N}_2 + \text{H}_3^+ \rightarrow \text{N}_2\text{H}^+ + \text{H}_2
\end{equation}
\begin{equation}
    \text{N}_2\text{H}^+ + \text{CO} \rightarrow  \text{HCO}^+ + \text{N}_2 
\end{equation}
\citep{bergin01,van2017}. These reactions show that N$_{2}$H$^{+}$ forms from N$_2$ gas and is destroyed by CO gas. Also, CO competes with N$_2$ to react with $\text{H}_3^+ $ and form HCO$^+$ ($\text{CO} + \text{H}_3^+ \rightarrow \text{HCO}^+ + \text{H}_2$). Therefore, N$_{2}$H$^{+}$ is abundant when CO gas is depleted and N$_2$ gas is abundant. In other words, it is abundant in the regions where the gas temperatures are below the freeze-out temperature of CO and above that of N$_2$, i.e.,  10 K $<$ T$_{\text{gas}}$ $<$18 K \citep{qi19}. In the regions where the temperature falls within this range, a constant N$_{2}$H$^{+}$ abundance ($\text{X}_{\text{N}_2{\text{H}^+}}$) is added. We explore the impact of a CO freeze-out temperature of 20~K in section \ref{sec:CO_freeze}.

As for DCO$^+$, its chemistry is more complex than N$_2$H$^+$ since it can be formed via multiple pathways
\begin{equation}{\label{eq:h2d+}}
    \text{H}_2\text{D}^{+} + \text{CO} \rightarrow   \text{DCO}^{+} + \text{H}_2 
\end{equation}

\begin{equation}{\label{eq:D}}
    \text{HCO}^{+} + \text{D} \rightarrow  \text{DCO}^{+} + \text{H}
\end{equation}

\begin{equation}{\label{eq:ch2d+}}
    \text{CH}_2\text{D}^{+} + \text{O} \rightarrow \text{DCO}^{+} + \text{H}_2
\end{equation}
\citep{Adams1985, wooten87, millar1989, oberg2015, favre2015}. Additional pathways involving products of $\text{CH}_2\text{D}^{+}$ also contribute to \dcoplus formation, e.g.,  $\text{CH}_4\text{D}^{+} + \text{ CO} \rightarrow \text{DCO}^{+} +  \text{ CH}_4$ \citep{favre2015,oberg2015,aikawa18}. The $\text{H}_2\text{D}^{+}$ pathway or cold pathway (Eq. \ref{eq:h2d+}, hereafter CP) is most efficient at low temperatures (T$_{\text{gas}}$ $<$ 30 K)  since the destruction of \htwodplus via \htwodplus + H$_2$ $\rightarrow$ H$_3^+$ + HD is non-negligible at T$_{\text{gas}}$  $\gtrsim$ 30 K \citep{wooten87,millar1989}. The D and $\text{CH}_2\text{D}^{+}$ pathways or warm pathways (Eq. \ref{eq:D} and Eq. \ref{eq:ch2d+}, hereafter WP) dominate the production of \dcoplus beyond 30 K and are efficient up to $\sim$ 70 K, above which the $\text{CH}_2\text{D}^{+}$ enhancement is reduced due to reactions with H$_2$
\citep{millar1989, favre2015}. Therefore, in our parametric framework, we associate each pathway with the following temperature ranges: for the CP, we use $18 \; \text{K}<\text{T}_{\text{gas}}  < 30  \; \text{K}$ and for the WP $30 \; \text{K}<\text{T}_{\text{gas}}  < 70  \; \text{K}$. We use 18 K as the lower limit for the CP since it corresponds to the CO freeze-out temperature, below which CO is strongly depleted from the gas phase and \dcoplus formation via the CP becomes inefficient. Similarly to \ntwohplusnospace, we explore a case with CO freeze-out at 20~K in section \ref{sec:CO_freeze}.

Within these temperature ranges, we assign the abundances  $X_{\text{DCO}^+, \; warm}$ and $X_{\text{DCO}^+, \; cold}$ to the warm and cold pathways respectively. To account for the decreasing efficiency of deuterium fractionation with increasing temperature, we adopt an abundance prescription with an exponential temperature dependence: 

\begin{equation}{\label{eq:xdco_warm}}
    X_{\mathrm{DCO}^+, \; warm} (T_{gas})= \mathrm{X}_{\mathrm{DCO}^+, w} \times  \exp \left [\frac{E_{w}}{T} - \frac{E_w}{{T_{ref,w}}}\right]
\end{equation}

\begin{equation}{\label{eq:xdco_cold}}
    X_{\mathrm{DCO}^+, \; cold} (T_{gas}) = \mathrm{X}_{\mathrm{DCO}^+,c} \times  \exp\!\left [\frac{E_c}{T} - \frac{E_c}{{T_{ref,c}}} \right]
\end{equation}

\noindent where $T_{ref,w}$ and $T_{ref,c}$ are reference temperatures for the WP and CP respectively. We set these values roughly to the midpoint of the corresponding temperature ranges, i.e., $T_{ref,w}$ = 50 K and $T_{ref,c}$ = 25 K. The parameters $\text{X}_{\text{DCO}^+,w}$ and $\text{X}_{\text{DCO}^+,c}$ therefore correspond to the \dcoplus abundances at  $T_{gas} = T_{ref,w}$  and $T_{gas} = T_{ref,c}$ respectively. $E_w$ and $E_c$ represent the zero-point energy differences associated with the CH$_3^+$ - CH$_2$D$^+$ and H$_3^+$ - H$_2$D$^+$ exchange reactions. We adopt $E_w$ = 370 K  and $E_c$ = 232 K \citep{smithadamsalge82,wooten87,millar1989}

\subsection{Model Parameters}
For our model exploration, we fix the parameters listed in Table \ref{tab:fixed_params}. We adopt the same values as \citet{Flaherty2020} for the surface density parameters (M$_{\text{gas}}$, R$_{\rm c}$, $\gamma$), the temperature structure ($q$, Z$_{q0}$, T$_{\text{atm0}}$, T$_{\text{mid0}}$), geometry (position angle and  inclination), stellar mass, systemic velocity, and distance.  We set the turbulence parameter, $\delta \text{v}_{\text{turb}}$, to zero as a limiting case since models predict that turbulence decreases towards the disk midplane \citep{Simon2015}. We explore the impact of a higher turbulence (30 $\%$ of the local sound speed) comparable to that inferred from CO \citep{Flaherty2020}. We find that increased turbulence leads to a modest enhancement in the outer disk (Appendix \ref{sec:turb}). The free parameters in our model exploration are therefore the \ntwohplus abundance ($\text{X}_{\text{N}_2{\text{H}^+}}$) and the abundances at 25 and 50 K for the CP and WP for \dcoplus ($\text{X}_{\text{DCO}^+, w}$ and $\text{X}_{\text{DCO}^+,c}$). 

\begin{deluxetable}{cc}[h!]
\tablecaption{Fixed Model Parameters\label{tab:fixed_params}}
\tablewidth{0pt}
\tablehead{\colhead{Parameter} & {Value}}
\decimals
\startdata
$\delta_{vturb}$ & 0\\
log$_{10}$(R$_{c}$) & 2.44 au [1] \\
T$_{\text{atm}0}$ & 24.68 K [1]\\
T$_{\text{mid}0}$ & 14.3 K [1]\\
Inclination (i) & 36$^{\circ}$ [1] \\
Position angle (PA) & 154.8$^{\circ}$ [1] \\
Temperature index ($q$) & -0.371 [1] \\
System Velocity (V$_{\text{sys}}$) & 6.017 km s$^{-1}$ [1] \\
M$_{\text{gas}}$ & 0.04 M$_\sun$ [1,2,3]\\
Z$_{q0}$ & 70 au [1,4]\\
M$_{\text{star}}$ & 0.54 M$_\sun$ [1,5]\\
Distance & 144.5 pc [1,6]\\
Surface density index ($\gamma$) & 1 [1,7]\\
\enddata
\tablecomments{(1) \citet{Flaherty2020}; (2) \citet{Andrews2011}; (3) \citet{mcclure16}; (4) \citet{Rose2013}; (5) \citet{Simon2000}, \citet{simon19} (6) \citet{Lindegren2018}, \citet{bailer18}; (7) \citet{Hartmann1998}}
\end{deluxetable}
\vspace{-5mm}

To determine the best-fit values for these abundances, we employ a downhill simplex (Nelder-Mead) optimization algorithm. The algorithm
iteratively explores parameter space by evaluating multiple trial abundance sets and computing the reduced $\chi^2$ between the observed and modeled visibilities for each trial. During each iteration, the algorithm selects new trial abundance sets based on previous evaluations, retaining those that improve the fit and discarding those that do not. This process continues until the variation in $\chi^2$ falls within the tolerance, indicating convergence toward the abundance values that minimize $\chi^2$. The resulting abundance values, along with the corresponding reduced $\chi^2$ and Bayesian Information Criteria (BIC) are summarized in Tables \ref{n2hp_table} and \ref{dcop_table}.

\section{Model Results}\label{sec:mod_results}
\subsection{Fiducial model}\label{sec:mod_fiducial}
The first row of Figure \ref{fig:radprofile_all} shows the model radial profiles for N$_2$H$^+$ and \dcoplus using the pathways described above, and assuming the model parameters derived from CO (Table~\ref{tab:fixed_params}). This model roughly matches the location of the peaks for both \ntwohplus and \dcoplusnospace, suggesting that the location of these features is consistent with the chemical pathways described above. However, it does not produce the extended emission beyond 200 au seen in both \ntwohplus and \dcoplusnospace. This discrepancy can also be seen in the channel maps, where the model emission is more compact than the data (Figure \ref{fig:velchan_all}). In the following subsections, we explore modifications to our model to reproduce the observed radial profiles. 

\subsection{Varying CO freeze-out temperature\label{sec:CO_freeze}} The locations of \ntwohplus and \dcoplus emission are sensitive to the temperature at which CO freezes out. We initially assume this happens at 18 K, but direct measurements of CO gas temperature find minima at $\sim$20 K, suggesting that CO freezes out at 20 K  \citep{pinte2018,law21,galloway25}. Studies of \ntwohplus find CO condensation temperatures of 15 -- 25 K among different sources including DM Tau \citep{qi19}. Lower CO condensation temperatures are associated with binding onto pure CO ice \citep{martin-domenech2014,fayolle2016}, while higher condensation temperatures are possible when binding onto a mixture of different ices  \citep{collings2003,noble2012}.

\begin{figure*}[ht!]
\center
\includegraphics[scale=0.6]{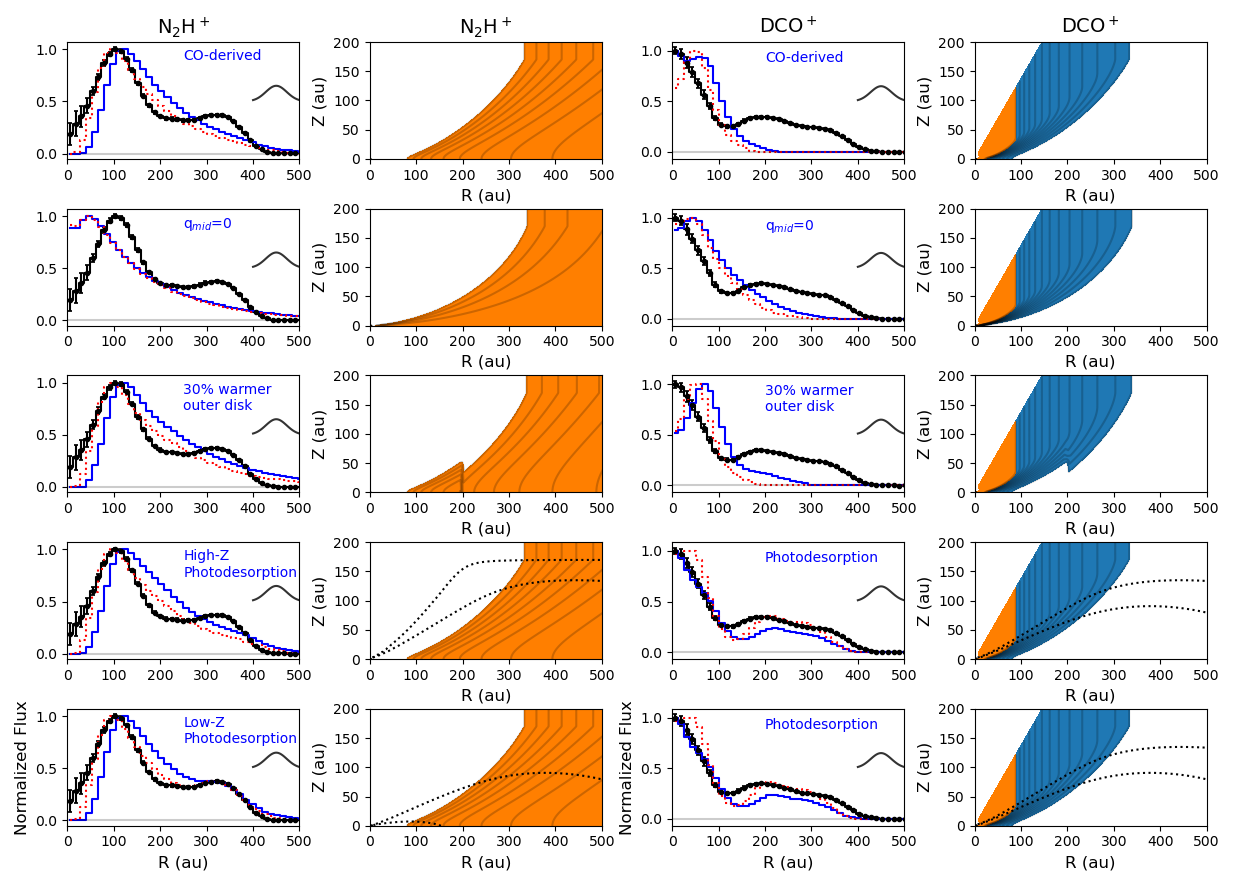}
\caption{Comparison of \ntwohplus and \dcoplus data with the models explored in this work. Each row corresponds to one of the models. From left to right, the first (\ntwohplusnospace) and third (\dcoplusnospace) panels show the radial profiles of the data (black),  while blue solid lines represent models with a CO freeze-out temperature of 18 K (fiducial value) and red-dashed lines show models with a freeze-out temperature of 20 K. The second panel shows the region between 10 K (N$_2$ freeze-out) and 18 K (CO freeze-out) where \ntwohplus is present in the model. The fourth panel highlights the regions with temperatures in the range relevant for \dcoplus formation, with the warm pathway dominating between 30 and 70 K (orange) and the cold pathway between 18 and 30 K (blue). The gray lines in panels 2 and 4 are temperature contours in steps of 1 K. \textit{Row 1:} Fiducial model generated with CO-derived parameters (Table \ref{tab:fixed_params}).  \textit{Row 2:} Model with a midplane temperature profile that is flat with radius ($q_{\rm mid}=0$).  \textit{Row 3:} Model with a midplane temperature profile that increases by 30\%\ beyond 200 au to mimic the effect of increased heating beyond the edge of  continuum emission \citep{Cleeves2016a}. \textit{Row 4:} Model including an \ntwohplus and \dcoplus ring in the outer disk to simulate the impact of increased photodesorption and ionization. For \ntwohplusnospace, the extended emission originates from the region above the CO photodesorption layer (0.01 $<\Sigma_{21}<$ 0.79, dashed lines in panel 2), while for \dcoplusnospace, the emission is enhanced in that layer (0.79 $< \Sigma_{21} <$ 3, dashed lines in panel 4). \textit{Row 5:} Same as row 4 except the \ntwohplus ring is closer to the midplane (3 $<$ $\Sigma_{21}$  $<$ 100).}
\label{fig:radprofile_all}
\end{figure*}

To explore the effect of the freeze-out temperature, we consider models in which freeze-out occurs at $T_{\rm gas}$ = 18~K (blue solid lines in Figure \ref{fig:radprofile_all}) as well as models with freeze-out at $T_{\rm gas}$ = 20~K (red dashed lines in Figure \ref{fig:radprofile_all}). The main effect is the change in radial location of the inner peak of the emission. For \ntwohplusnospace, the inner peak in the fiducial model moves inwards by $\sim$20 au for a larger freeze-out temperature, and a similar effect is seen throughout many of the models, as well as in \dcoplusnospace, except the model with a constant midplane temperature (discussed below).

We note that there is a degeneracy between the CO freeze-out temperature and the gas temperature in the disk, as a warmer disk with a larger CO freeze-out temperature will have a CO snowline at the same location as a colder disk with a lower CO freeze-out temperature. The utility of \ntwohplus and \dcoplus as tracers of the CO snowline is discussed in section \ref{sec:snowline}.

\subsection{Warm Midplane}\label{mod_warmmidplane}
The mismatch between the data and the models in the outer disk may be explained by a higher midplane temperature than assumed in the fiducial model. With a higher gas temperature, the \ntwohplus abundance may decrease at intermediate radii as CO is returned to the gas phase, while increasing at larger radii as the N$_2$ condensation front is pushed outwards. The return of CO to the gas phase will increase the abundance of \dcoplus produced through the cold pathway. An increase in the temperature will also directly increase the intensity, independent of any change in the abundance.  

We first consider a model in which the midplane has a flatter temperature profile than the surface layers (i.e., $T_{\rm mid}\sim r^{q_{\rm mid}}$, $T_{\rm atm}\sim r^{q_{\rm atm}}$, with  $|q_{\rm mid}|<|q_{\rm atm}|$). Predicted by some radiative transfer models \citep[e.g.,][]{D'Alessio2006}, this will result in a higher midplane temperature than in the fiducial model. We explore $q_{\rm mid}$ = 0 as a limiting case. We find that the temperature increase in the outer disk midplane raises modestly the flux of \ntwohplus and \dcoplus emission, but it does not create the rings seen in the data  (second row of Figure~\ref{fig:radprofile_all}). The outer disk temperature does not rise above the CO freeze-out temperature, thus the increased emission is entirely explained by the relationship between intensity and temperature. 

Next, we consider a sharp increase in the midplane gas temperature beyond the outer edge of the mm-disk. The chemical models of \citet{Cleeves2016a} show that the radial drift of dust grains reduces the opacity in the outer disk, resulting in a temperature increase that can range between 10 and 30\%. The dust emission around DM Tau is more compact ($\sim$ 200 au)  than the  \ntwohplusnospace, \dcoplusnospace, and CO isotopologue emission (r $>$ 400 au; Figure \ref{fig:n2h_dco_co_cont_rad_profs}), which is suggestive of radial drift that can alter the thermal structure in the outer disk. To simulate this effect, we artificially increase the temperature beyond 200 au, the edge of the mm continuum emission (Figure \ref{fig:n2h_dco_co_cont_rad_profs}), by 30\% above its fiducial value, the maximum change reported by \citet{Cleeves2016a}.  This results in a modest increase in \ntwohplus and \dcoplus emission beyond the continuum edge, but it does not generate the outer rings (third row of Figure \ref{fig:radprofile_all}). This mismatch suggests that a thermal inversion alone is not enough to reproduce the extended emission.

\begin{figure}[t!]
\center
      \includegraphics[width=0.35\textwidth]{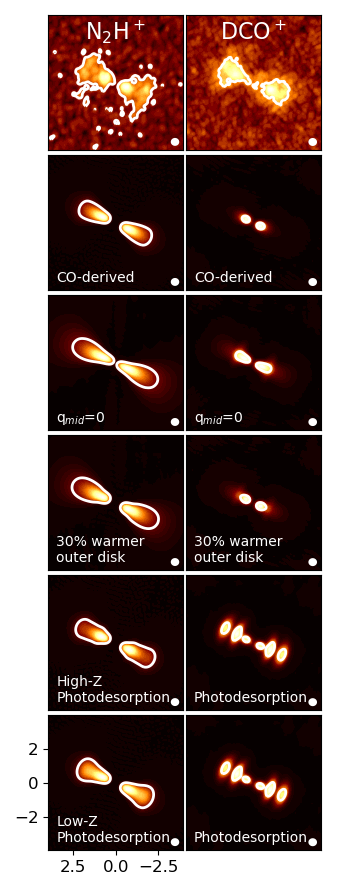}
\caption{Central velocity channels for \ntwohplus (\textit{left}) and \dcoplus (\textit{right}) data and models. The beam sizes are shown in the bottom right corner of each panel and the white contours are 3$\sigma$ contours. \textit{Row 1:} Data. \textit{Row 2:} Fiducial model based on CO-derived parameters. \textit{Row 3:} Model with a flat midplane temperature profile ($q_{\rm mid}=0$). \textit{Row 4:} Model with a 30 $\%$ temperature jump at radii beyond 200 au. \textit{Row 5:} Model including \dcoplus and \ntwohplus rings in the outer disk to mimic increased photodesorption and ionization. The \dcoplus ring is placed in the CO photodesorption layer, while the \ntwohplus ring is placed above that layer. \textit{Row 6} same as row 5 expect the \ntwohplusnospace ring is closer to the disk midplane.}
\label{fig:velchan_all}
\end{figure}
\vspace{2mm}
\subsection{Photodesorption and ionization in the outer disk\label{sec:photod_models}}
Another process is needed to explain the emission observed in the outer disk. At these large radii, the surface density of the disk starts to decrease rapidly, which can allow for greater penetration of UV photons. Radial drift can also further enhance the penetration depth \citep{Cleeves2016a}. This increased radiation can trigger the photodesorption of CO and N$_2$, the precursors to \dcoplus and \ntwohplus respectively, which would otherwise remain frozen in the cold outer disk. In addition, the reduced opacity can enhance cosmic ray and X-ray ionization rates, which also impact abundances.  The formation of \ntwohplus (Eq. \ref{eq:n2h_formation}) and cold \dcoplus (Eq. \ref{eq:h2d+}) require H$_3$$^+$, the precursor to \htwodplusnospace. Since H$_3^+$ is produced through the ionization of H$_2$ \citep{carney18}, an enhanced ionization rate can contribute to the extended emission. As a result of the increased UV radiation and cosmic ray and X-ray ionization rates, extended emission and ring-like features can appear at large radii. For example, a secondary \dcoplus ring has been observed in IM Lup and attributed to UV photodesorption \citep{oberg2015}. Similarly, chemical models of DM Tau that include interstellar UV photons show enhanced \dcoplus emission compared to models without these photons \citep{teague15}. 

To approximate the effect of these processes within our parametric framework, we include an additional ring component for both \ntwohplus and \dcoplus  in the outer disk. The radial width of this ring is set to 50 au, while its inner radius is treated as an additional free parameter in the downhill simplex fitting procedure. For \dcoplusnospace, whose cold formation requires the presence of CO gas, we set the vertical extent of this ring based on the expected location for CO photodesorption. The return of CO to the gas phase is mediated by the absorption of UV photons and is balanced by the rate at which CO freezes out.  Previous analysis \citep{Flaherty2020} indicates that this effect can be approximated by placing CO between lines of constant surface density. Thus, for \dcoplusnospace, we use the photodissociation boundaries $0.79 \leq \Sigma_{21} \leq 3$ where $\Sigma_{21}$ = $\rm{N}_{\rm{H}_2}$/(1.59 $\times 10^{21}$ cm$^{-2}$) \citep{Flaherty2017,Flaherty2020}. 

As for \ntwohplusnospace, it can form below the CO photodesorption layer where CO remains frozen out, N$_2$ is still in the gas phase, and the ionization rates enhances H$_3^+$ production close to the midplane. In addition, models of \citet{van2017} show that \ntwohplus can exist in a layer well above the midplane due to CO photodissociation and N$_2$ shielding. Similarly, the models of \citet{aikawa15} reveal a layer of \ntwohplus at z/r$\sim$0.2, where CO has been converted into CO$_2$, while photodissociation prevents N$_2$ from being reprocessed into NH$_3$ ice.  Thus, \ntwohplus may be present at large radii (1) above the CO photodesorption layer, due to CO photodissociation and conversion of CO into CO$_2$, and/or (2) deeper than the photodesorption layer, due to N$_2$ photodesorption and increased ionization. To simulate these scenarios, we use the following photodissociation boundaries for \ntwohplusnospace: $ 3 < \Sigma_{21} < 100$ for the photodesorption and increased ionization scenario and $0.01 < \Sigma_{21} < 0.79$ for the CO photodissociation and CO--CO$_2$ scenarios. 

\begin{figure*}[thb!]
    \center
    \includegraphics[scale=.37]{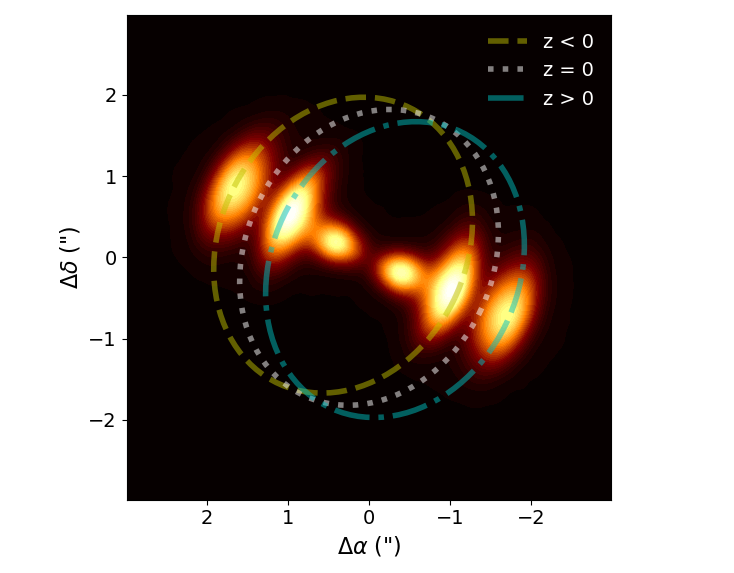}
    \includegraphics[scale=.4]{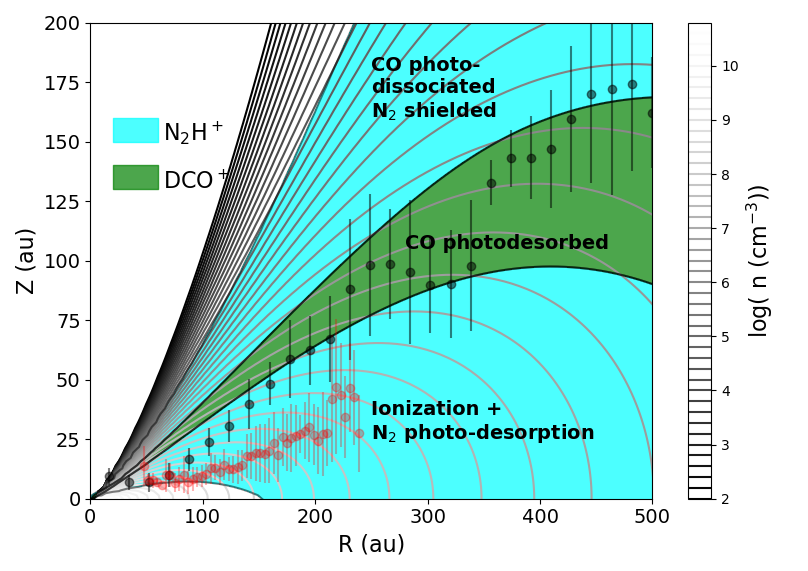}
    \caption{\textit{Left:} \dcoplus model image with emission high in the disk. Dashed lines are a ring at r = 220 au and z = -30 (yellow), 0 (white), +30 au (cyan). In this model image, the two bright features in the outer disk are the emission from above and below the midplane.  \textit{Right:} Column density boundaries of $\Sigma_{21}$ = [0.01, 0.79, 3, 100], showing the approximate regions where different physical mechanisms for supplying \ntwohplus and \dcoplus in the outer disk may operate. Black and red circles represent the height of the emitting regions for $^{12}$CO and $^{13}$CO respectively, as derived by \citet{galloway25}. The predicted region for CO photodesorption in the outer disk is consistent with the observed emitting region for $^{12}$CO. \label{fig:photod}}
\end{figure*}

Placing \ntwohplus and \dcoplus high in the disk leads to, what appears to be, two distinct rings in the outer disk (Figure \ref{fig:velchan_all}). This effect is more noticeable for \dcoplusnospace, while for \ntwohplusnospace, the ring emission blends with the inner disk.  This multiple-ring effect is due to the deprojection used to generate the radial profile, which assumes that the emission is coming from $z=0$. When the emission comes from the surface layers, the $+z$ and $-z$ layers are spatially distinct (Figure ~\ref{fig:photod}), and appear at two different deprojected radii, an effect that is often seen with CO \citep{law2022,law2023}. In the case of \ntwohplusnospace, this allows us to rule out its presence high in the disk, as the data do not show this behavior. In the model with \ntwohplus closer to the midplane (Figure \ref{fig:velchan_all}), the double-ring effect is no longer present. In addition, the radial profile for this model better matches the \ntwohplus morphology, especially between $\sim$ 200 and 400 au (Figure \ref{fig:radprofile_all}).

In contrast, for \dcoplusnospace, the model with the high outer ring reproduces the overall shape of the observed radial profile, although the extended emission remains slightly fainter than the data for a CO freeze-out temperature of 18 K and the outer rings do not blend as smoothly with the inner disk (Figures \ref{fig:radprofile_all} and \ref{fig:velchan_all}). Nevertheless, among the processes explored in this work, photodesorption and ionization are the mechanisms that most capture the spatial extent of the extended emission and the observed outer disk feature, especially for \dcoplusnospace.  This overall agreement raises the possibility that the extended \dcoplus structure could be explained by CO photodesorption high in the atmosphere, without the need to invoke a temperature enhancement at the midplane. However, CO photodesorption and a temperature enhancement are degenerate in their effects on \dcoplusnospace, in the sense that a higher temperature could compensate for a lower \dcoplus abundance in the outer ring. Disentangling this degeneracy requires independent constraints on the \dcoplus density and temperature (e.g., \dcoplus line ratios). 

Assuming that the \dcoplus emission beyond 100 au represents two rings at the same radial distance, but at different heights above the midplane, rather than two concentric rings in the midplane at different radial distances, we can estimate the height ($z$) above the midplane for the emission based on the projected distance of the rings ($r_1$, $r_2$) and the known inclination ($i$) of the disk as,

\begin{equation}
    z = \frac{r_2-r_1}{2\tan i}
\end{equation}
The observed \dcoplus emission peaks at $r_1\sim190$ au and $r_2\sim320$ au imply a pair of rings at a radial distance of $r = (r_1+r_2)/2=$ 255 au and $z=$ 90 au above and below the midplane. This location places \dcoplus within the expected CO photodesorption region around DM Tau (Figure ~\ref{fig:photod}). It is also consistent with the vertical location of CO inferred by \citet{law2023} and \citet{galloway25}. Further observations are needed to confirm the vertical origin of this ring, and to better constrain its radial profile. In addition to the comparison of the radial profiles and central velocity channels, the BIC values in Tables \ref{tab:n2hmodel_params} and \ref{tab:dcomodel_params} further suggest that the models accounting for photodesorption and ionization in the outer disk are favored, as they yield lower BIC values. 

\begin{figure*}[t!]
    \center
\includegraphics[scale=0.6]{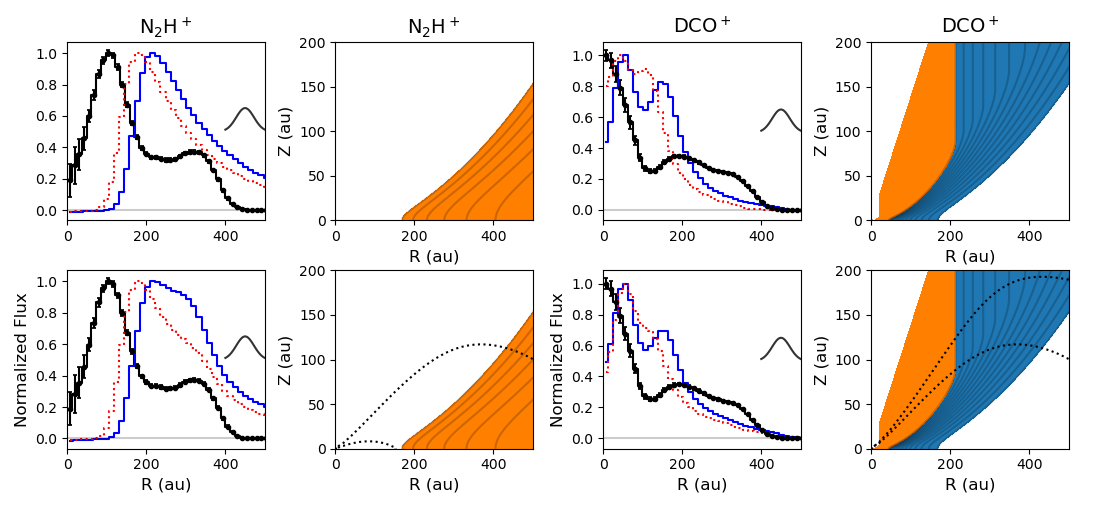}
\caption{Similar to Figure~\ref{fig:radprofile_all}, except the  models were generated using the temperature structure derived by \citet{galloway25}. The top row shows the fiducial models (Section \ref{sec:mod_fiducial}), while the bottom panel includes additional emission in the outer disk (Section \ref{sec:photod_models}). The warmer gas temperatures derived by \citet{galloway25} push the predicted \ntwohplus and \dcoplus emission to larger radii, but are still unable to reproduce the extended emission morphology seen in both lines.}
\label{fig:radprofile_gs25}
\end{figure*}

\subsection{Alternative temperature structures}\label{sec:alt_temp}
Our parametric model suggests that the extended emission is driven by increased UV penetration and cosmic ray/X-ray ionization rates in the outer disk. While we have explored alterations to our temperature structure (Section \ref{mod_warmmidplane}), it is possible that a different temperature parametrization might better capture the emission without invoking additional mechanisms. Thus, we discuss some of the temperature profiles in the literature and test our fiducial and best-fit models with a different profile.   Various studies have attempted to reconstruct the temperature structure of the disk around DM Tau. Some have employed radiative transfer models to calculate dust/gas temperatures while accounting for sources of irradiation \citep{loomis2015,teague15,qi19,zhang2019,hardiman2026}. Other have used simple parametric models to fit observations of molecular emission lines \citep{Dartois2003,pietu2007,semenov2018,Flaherty2020}. More recent studies have used high-resolution observations to directly measure the height and radius of molecular line emission, and in turn used the brightness temperature of these optically thick emission lines to directly constrain the gas temperature in the disk \citep{law2022,law2023,galloway25}. These studies find similar, but not identical, temperature structures, reflecting the inherent uncertainty in estimating the temperature structure. In general, earlier studies \citep[e.g.][]{teague15} derive a colder midplane and a warmer surface layer, which we can rule out as it places the midplane CO snowline at $\sim$10 au.

Recently \citet[][hereafter GS25]{galloway25} used $^{12}$CO($J=3-2$), $^{13}$CO($J=3-2$), and CS($J=7-6$) observations of the disk around DM Tau taken as part of the exoALMA survey \citep{teague2025} to derive the emitting surfaces of these molecules, the gas temperatures at these emitting surfaces, and consequently the two-dimensional temperature structure. The median temperature structure derived by GS25 is warmer than that derived by \citet{Flaherty2020} and a warmer disk may produce \ntwohplus and \dcoplus emission at larger radii as the CO and N$_2$ snowlines are pushed to larger radii.

GS25 uses the same functional form as used here, but with different values for the temperature normalization. Specifically, they derive $T_{\rm atm0}=34.2_{-4.1}^{+0.4}$ K, $T_{\rm mid0}=18.8_{-4.6}^{+0.2}$ K, $q_{\rm mid}=-0.5_{-0.1}^{+0.015}$, $q_{\rm atm}$=-0.13$_{-0.1}^{+0.015}$, and $Z_{q0}=56.3$ au, after re-normalizing $T_{\rm atm0}$, $T_{\rm mid0}$, and $Z_{q0}$ to a radius of 150 au. With this modified temperature structure, we run again the downhill simplex optimization algorithm to find the best \ntwohplus and \dcoplus abundances, considering the fiducial model (Section \ref{sec:mod_fiducial}) and a model with an extra ring in the outer disk (Section \ref{sec:photod_models}). The resulting radial profiles are shown in Figure~\ref{fig:radprofile_gs25}. We find that the warmer disk does lead to enhanced emission in the outer disk, but in a way that is inconsistent with the observations. The predicted \ntwohplus emission peaks at 200 au, instead of 100 au as observed in the data. The \dcoplus warm and cold pathways peak at $\sim$ 50 and 150 au, near the observed local minimum in \dcoplus emission at 120 au. Including additional emission in the outer disk has little effect on the final result as the model cannot account for the inner disk emission. 

Accurately constraining the midplane temperature within a protoplanetary disk is challenging, and the temperature is often inferred indirectly. \citet{Flaherty2020} analyze CO emission, which comes from well above the midplane, and adjust the $T_{\rm mid0}$ in their models so as to ensure $T_{\rm mid}$=19 K at r = 70 au, matching the chemical modeling of \citet{loomis2015}, which are generated to reproduce the H$_2$CO emission around DM Tau. GS25 use $^{13}$CO, which emits closer to the midplane, to better infer the temperature near the midplane. In both approaches, there are uncertainties in the derived temperature structure, and when accounting for uncertainties, the midplane temperatures from \citet{Flaherty2020} and GS25 are consistent with each other. It is only when using the median values from the posterior distributions from GS25 that differences arise. The difference in emission profiles when using either temperature structure speaks to the sensitivity of \ntwohplus and \dcoplus to the midplane temperature; a difference of a few Kelvin is enough to move the location of the \ntwohplus and \dcoplus emission by many tens of au. With this comparison, we find that while hotter temperatures can enhance the \ntwohplus and \dcoplus emission, they do not reproduce well the morphology of the extended emission. GS25 also reported extended CO emission in the DM Tau disk and three other disks, which they attribute to photodesorption. 

We can use line ratios for \ntwohplus and \dcoplus to constrain disk-averaged excitation temperatures for these molecules. \citet{Long24} examine \ntwohplus ($J=4-3$) and ($J=3-2$), constraining an excitation temperature of 25 K at 100 au. Using \texttt{RADEX} \citep{vandertak17}, IRAM observations of \dcoplus ($J=2-1$) \citep{guilloteau2006} and SMA observations of \dcoplus ($J=3-2$) emission \citep{oberg2011b}, along with our ALMA \dcoplus ($J=4-3$) flux measurement (1.64 $\pm$ 0.2 Jy km s$^{-1}$), we constrain the gas temperature to $>$18 K. Using just the \dcoplus SMA ($J=3-2$) and ALMA ($J=4-3$) observations, the estimated excitation temperature is $>$8 K, suggesting the presence of faint cold \dcoplus emission that may not have been detected by the IRAM observations.

\section{Discussion} \label{sec:disc}
 \subsection{Extended emission in protoplanetary disks}
Our parametric model exploration shows that the extended \ntwohplus and \dcoplus emission beyond the dust disk around DM Tau is best explained by UV-driven photodesorption and increased cosmic ray and X-ray ionization rates, with a potentially modest contribution from a thermal inversion beyond the continuum edge. With our models, we are unable to disentangle the relative importance of photodesorption versus ionization. Recently, \citet{Long24} used a combination of \ntwohplus and HCO$^+$ observations from ALMA and a chemical code to characterize the ionization profile of the DM Tau disk. They find that the disk is best described by a radially decreasing ionization rate, with a sharp transition at 200 au. In other words, the inner disk is best fit by their high ionization models ($\geq$ 10$^{-17}$ s$^{-1}$), while the outer disk requires lower ionization models ($\sim$ 10$^{-18}$ s$^{-1}$). However, these low ionization models predict  fluxes fainter than observed in the outer disk, while higher ionization models produce emission brighter than observed.

The models of \citet{Long24} do not include external UV radiation, thus it is possible that UV-driven photodesorption could account for the missing emission in the outer disk that their ionization models cannot capture. In an attempt to constrain the relative contribution of UV-driven photodesorption vs. cosmic ray/X-ray ionization, we incorporate external UV radiation into their best-fit chemical model and re-run their chemical code.  Their chemical models are well suited for this investigation since they adopt the temperature structure from \citet{Flaherty2020}, which we also use in this work, allowing us to maintain a consistent thermal structure. We test UV strengths of 1 G$_0$, 5 G$_0$, and 10 G$_0$ where G$_0$ = $1.6\times10^{-3}$~erg~cm$^{-2}$~s$^{-1}$ corresponds to the average interstellar radiation field \citep{Habing1968}. Their code does not include deuterated species, thus we focus on the effects of the UV radiation on \ntwohplus and the species affecting its abundance. 

\begin{figure*}[t!]
    \center
\includegraphics[scale=0.37]{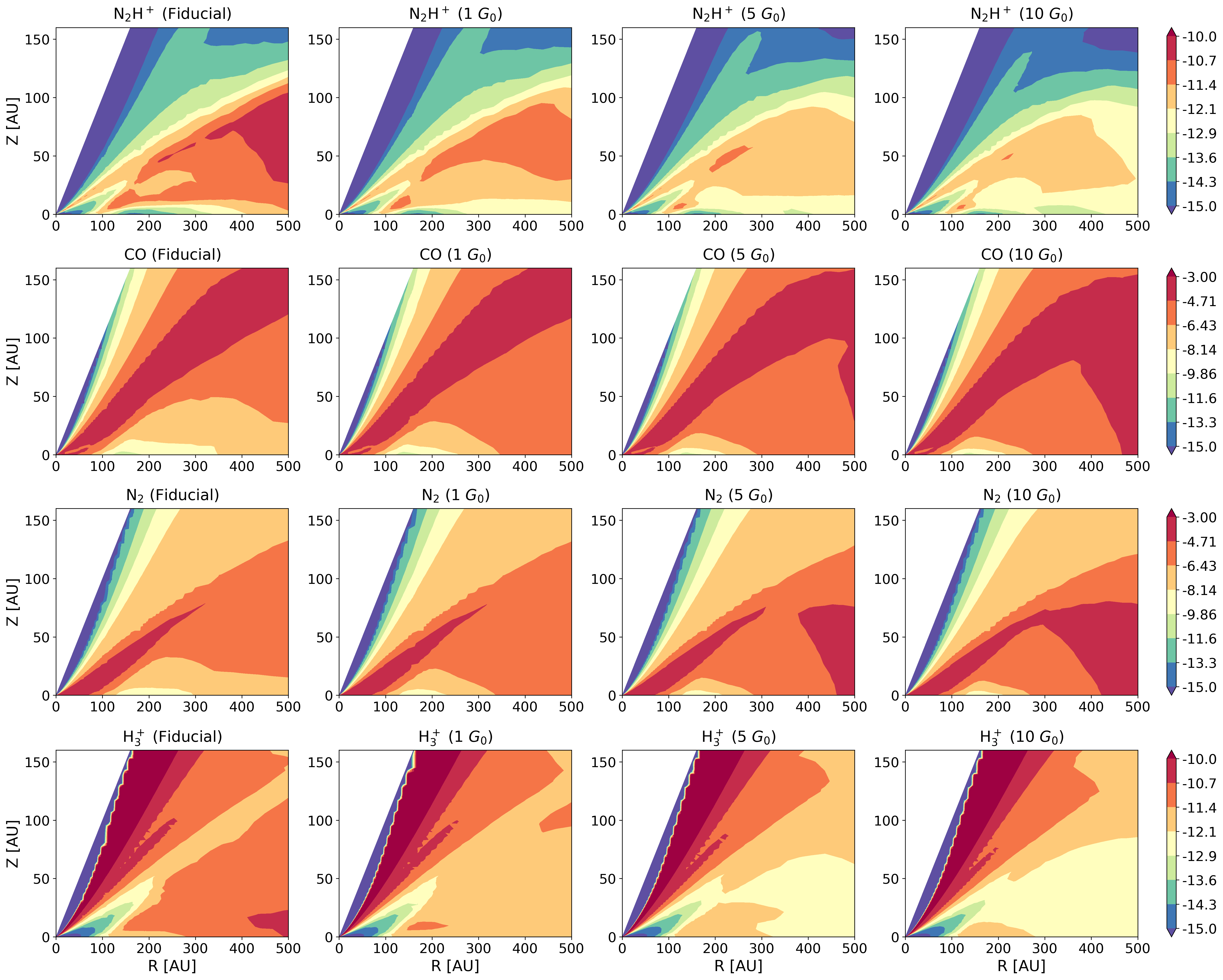}
\caption{Abundance distributions of $\mathrm{N_2H^+}$, CO, $\mathrm{N_2}$, and $\mathrm{H_3^+}$ in the DM Tau disk as a function of external UV radiation strength. We use the best-fit chemical model from \citet{Long24}, denoted as fiducial, and incorporate different amounts of external UV radiation.  From left to right, the columns show models with no external UV (fiducial) and UV strengths of 1, 5, and 10 G$_0$, respectively. Color bars indicate the abundance with respect to hydrogen on a logarithmic scale, $\log_{10}(n_{\rm X}/n_{\rm H})$.}
\label{fig:chem_models}
\end{figure*}

Figure \ref{fig:chem_models} shows \ntwohplusnospace, CO, N$_2$, and H$_3^+$ abundances as a function of the external UV field strength. As the UV radiation increases, the \ntwohplus abundance decreases, while the CO and N$_2$ abundances increase as a result of photodesorption. The reduced \ntwohplus abundance can be attributed to the decrease of its precursor,  H$_3^+$, as a result of increasing UV. Thus, while an enhanced UV field increases the abundance of CO and N$_2$, it decreases the abundance of H$_3^+$. This implies that the extended emission in the DM Tau disk can likely be reproduced with a higher level of  cosmic ray/X-ray ionization than previously reported, which maintains high H$_3^+$ abundance combined with increased photodesorption that brings CO and N$_2$ back to the gas phase. In other words, the extended emission arises from a balance between cosmic ray/X-ray ionization and external UV. Quantifying the contribution of each mechanism requires more extensive chemical modeling and a multi-molecular analysis including UV tracers such as C$_2$H, CN, and NO, which is beyond the scope of this work.

There is evidence that the outer regions of the DM Tau disk are exposed to some amount of external UV radiation. Indeed, \citet{Bergin16} report the detection of a C$_2$H ring at $\sim$ 350 au, which their thermochemical model can only reproduce when there is enhanced UV radiation down to the outer disk midplane and a C/O ratio $>$ 1. Moreover, \citet{galloway25} attribute the observed diffuse CO emission at large radii around DM Tau to CO photodesorption. \citet{rossella25} estimated the UV flux in various star-forming regions including Taurus where DM Tau belongs. They report a median external UV of 2.92$^{+0.16}_{-0.14}$ G$_0$, showing that some level of external UV radiation can impact disks in Taurus.  Taken together, our parametric and chemical modeling supports external UV photodesorption and enhanced cosmic-ray/X-ray ionization as the  dominant drivers of the extended (r $>$ 200 au) molecular emission. A thermal inversion can still contribute to the emission at the midplane, but  previous models of the temperature structure around DM Tau have not found evidence for such an inversion. It is important to note that these models derived the gas temperature using dust radiative transfer \citep[e.g.,][]{teague15,zhang2019,hardiman2026}, and extrapolating into dust-free regions, or employed observational tracers that are not sensitive to the midplane beyond the continuum outer edge \citep[e.g.,][]{law2023,galloway25}.

In addition to the processes discussed above, other factors may influence the observed emission structure. The \dcoplus abundance is also sensitive to the ortho-to-para (o/p) ratio of H$_2$. \htwodplusnospace, the precursor to \dcoplusnospace, is destroyed by H$_2$. This reaction is more efficient for ortho-H$_2$, whose internal energy (170.5 K) effectively reduces the energy barrier (232 K) required for the reaction to proceed \citep{walmsley04,murillo15,aikawa18}. As a result, an elevated H$_2$ o/p ratio between 0.1 and the statistical limit of 3 can suppress cold \dcoplus production by increasing the destruction rate of \htwodplusnospace \citep{teague15}. Local variations in this ratio may thus imprint on the spatial distribution of \dcoplusnospace. In addition, while our analysis focuses on chemical and thermal processes, we cannot rule out the possibility that radial variations in the gas density may also play a role. Some pressure-gradient features inferred from CO kinematics are located near the \ntwohplus peak and \dcoplus depression at $\sim$ 100 au \citep{stadler25}, and such features may contribute to shaping the morphology within 100 au.
 
The presence of extended emission out to hundreds of au is not unique to the DM Tau system. A particularly similar case is the IM Lup disk where two concentric rings of \dcoplus have been observed: an inner ring consistent with a radial temperature decrease and an outer ring attributed to UV-driven photodesorption in regions where CO would otherwise be frozen out \citep{oberg2015,Cleeves16b}. Furthermore, extended \dcoplus and \ntwohplus emission has been reported around V4046 Sgr, HD163296, HD 169142, GM Aur, LkCA 15, TW Hya, GG Tau A \citep{huang17,salinas17,Flaherty2017,carney18,qi19,munozromero,Kashyap24}. Similar outer disk emission has also been found in other molecular tracers. For instance, \citet{galloway25} report CO emission extending out to large radii around AA Tau, LkCa 15, and SY Cha in addition to DM Tau.  Extended C$_2$H, N$_2$D$^+$, DCN, HCN, and H$_2$CO emission has also been observed \citep[e.g.,][]{Bergin16,Bergner19,cataldi21,guzman21,law21}. 

In many cases, this emission comes from regions beyond the continuum edge, suggesting that, similar to DM Tau, UV-driven photodesorption, cosmic ray/X-ray ionization, and potentially thermal inversion play an important role in shaping outer disk chemistry. The importance of each process can vary across disks, depending on factors such as the radial and vertical structures, gas mass, and the presence or absence of substructures. Quantifying the contribution of each mechanism will require high-resolution, multi-molecular observations. For example, species like C$_2$H, CN, and NO can trace the UV radiation field \citep[e.g.,][]{leemker24}; multi-CO line studies, line ratios, and radial excitation analysis can provide insights into midplane temperatures and potential thermal inversions \citep[e.g.,][]{schwarz2019,calahan2021,munozromero}; and species such as N$_2$D$^+$ can further constrain ionization levels in the outer disk near the midplane \citep[e.g.,][]{aikawa2021}.

\subsection{\ntwohplus and \dcoplus as CO snowline tracers\label{sec:snowline}}
Snowlines mark radial locations in the disk midplane where volatiles, such as H$_2$O, CO$_2$, and CO, begin to freeze out onto dust grains due to the radial drop in gas temperatures. Constraining the locations of these snowlines is important since (1) they influence the evolution of dust grains, and thus the formation of rocky planets and cores, and (2) they regulate the distribution of gas-phase versus solid-phase C/N/O ratios, which affect the composition of nascent planets \citep[][and references therein]{oberg11,qi19,oberg2023}. The CO snowline is the most accessible observationally since it is located farther from the star as a result of the lower freeze-out temperature of CO \citep{oberg11,qi13,van2017}, and is essential for the formation of many complex organic molecules as they require CO ice \citep{Herbst09}. However, directly locating the midplane CO snowline has been a challenging endeavor since CO is optically thick and traces the disk surface \citep{Dartois2003,van2017,law2022,law2023}. 

Alternatively, \ntwohplus and \dcoplus have been proposed as indirect tracers of the CO snowline given that their chemistry is tied to CO. For \ntwohplusnospace, its inner radius is expected to coincide with the snowline \citep{qi13} since it is readily destroyed by CO gas \citep{bergin01,jorgensen04}. However, good knowledge of the temperature structure is needed since \ntwohplus can also be produced above the midplane \citep{aikawa15,nomura16,van2017}. The relationship between \dcoplus emission and the CO snowline is more complex due to the different formation pathways. If \dcoplus is predominantly formed via the CP, its radial distribution is expected to peak just interior to the CO snowline \citep{mathews13,qi15}. This behavior arises because H$_3^+$, the precursor to \htwodplusnospace, is destroyed by CO gas to form HCO$^+$, while deuteration via \htwodplus requires cold temperatures and some amount of CO gas to proceed. As a result, CO gas must be sufficiently abundant to enable cold \dcoplus formation, but not so abundant that it quenches \htwodplus production \citep{carney18,munozromero}. These conditions are satisfied in a relatively narrow temperature range near the CO snowline, therefore the \dcoplus peak could in principle be used to locate the snowline \citep{mathews13}. However, the reliability of this approach is compromised by the WP, which can produce \dcoplus in the inner disk and the disk surface \citep{favre2015,oberg2015,qi15}.  

In the case of DM Tau, we place the observed \ntwohplus and \dcoplus morphologies in the context of previous constraints on the CO snowline location. Accounting for the temperature structure, \citet{qi19} report a CO snowline located at $75^{+10}_{-30}$ au, constrained from visibility-domain modeling of the \ntwohplus ring inner edge. In our data, the \ntwohplus intensity peaks near $\sim$ 100 au, while \dcoplus shows a depression near 120 au. Given that \ntwohplus is readily destroyed by CO gas and that cold \dcoplus formation requires some CO in the gas-phase, these features are suggestive of a snowline closer to the higher-end estimate of \citet{qi19}. At the same time, both molecules show detectable emission down to $\sim$35 au (Section \ref{sec:obs_res}), which is within the broad lower-bound range reported by \citet{qi19} though such inner-disk emission could instead be driven by processes not directly tied to CO freeze-out.

For \dcoplusnospace, our models show that the inner disk emission includes a contribution from the warm pathway (Section \ref{sec:tt_herbig}). As for \ntwohplusnospace, it may be influenced by CO depletion  and elevated ionization rates in the inner disk. CO is depleted by a factor of 10 from its ISM abundance (CO/H$_2$ $\sim 10^{-4}$) in the inner 100 au \citep{zhang2019}, which can allow \ntwohplus to persist interior to the CO snowline. In addition, \citet{Long24} infer high ionization rates in the inner disk ($> 10^{-17}$ s$^{-1}$), which can enhance \ntwohplus production even in the presence of gas-phase CO \citep[e.g.,][]{cleeves14}. Furthermore, high ionization rates can enhance \ntwohplus emission by facilitating the chemical processing of CO into less volatile species \citep[e.g.,][]{Schwarz18,schwarz19}. More work is still needed to understand the exact mechanism causing the elevated ionization rates in the inner regions of the DM Tau disk \citep{Long24}.

In summary, these observed features are  overall compatible with the snowline constraints. However, the inner disk emission can be influenced by processes unrelated to CO freeze-out, as discussed above. Therefore, using \dcoplus as a CO snowline tracer requires disentangling the relative contributions of the warm and cold formation pathways, while for \ntwohplusnospace, the impact of ionization, CO depletion, and CO freeze-out on the inner edge of the emission must be further constrained.  Breaking these degeneracies can be achieved through complementary molecular tracers such as HCO$^+$ and N$_2$D$^+$ \citep[e.g.,][]{aikawa18,cataldi21}, as well as detailed chemical modeling that accounts for these processes in a self-consistent manner \citep[e.g.,][]{van2017}.

\subsection{\dcoplus pathways in T Tauri vs. Herbig disks}\label{sec:tt_herbig}
Previous observations of \dcoplus in protoplanetary disks have generally shown that the emission is dominated  by the cold pathway \citep[e.g.,][]{oberg2015,Flaherty2017,huang17,munozromero}. The warm pathway is usually negligible except for Herbig disks, e.g., HD 163296 and HD 169142, where its contribution needs to be taken into account to accurately reproduce observations \citep{salinas17, carney18,salinas18}. In addition, \citet{huang17} note that the two Herbig disks in their disk survey, MWC 480 and HD 163296, exhibit \dcoplusnospace/HCO$^+$ ratios similar to those measured in T Tauri disks. This is surprising at first glance, since the higher temperatures in Herbig disks are expected to hinder deuterium fractionation. They suggest that the WP could potentially explain the derived ratios. Taken together, these studies indicate that the WP may play a more significant role in Herbig disks than in T Tauri disks, consistent with the warmer temperatures in the former.  

\begin{figure}[htb!]
      \begin{center}
      \includegraphics[width=0.45\textwidth]{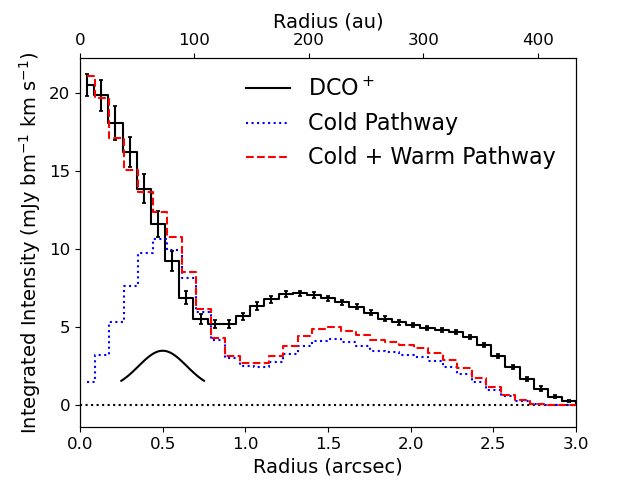}
      \end{center}
\caption{\dcoplus radial profiles for  the data (black), a model including the warm and cold formation pathways of \dcoplus (red), and a model including only the cold pathway (blue). The warm pathway is responsible for the peak and the inner disk emission within $\sim$ 70 au. }
\label{fig:DCO_coldVwarm}
\end{figure}

In the case of DM Tau, we re-ran our best-fit model without the inclusion of the warm pathway in order to assess its contribution to the observed \dcoplus emission. The resulting model indicates that the warm pathway is responsible for the peak and the inner disk emission within $\sim$ 70 au (Figure \ref{fig:DCO_coldVwarm}). This suggests that for DM Tau, the WP plays a non-negligible role in shaping the \dcoplus emission, making it stand out from previous T Tauri disks. More data sets comparing T Tauri and Herbig disks are needed to determine the relative importance of the different formation pathways for \dcoplusnospace.

\section{Conclusions} \label{sec:con}
In this work, we have examined the \ntwohplus and \dcoplus emission in the DM Tau disk to characterize the region near the midplane. \dcoplus is centrally peaked, while \ntwohplus peaks near 100 au. Both molecules also display extended emission out to 400 au. We investigate the origin of these features using parametric models of the disk density and temperature structure. The \ntwohplus peak is consistent with formation between the N$_2$ and CO snowlines, with the precise location sensitive to the adopted CO freeze-out temperature($\sim$ 18–20 K). For \dcoplusnospace, the central peak arises from its formation via the warm pathway, which dominates at 30 K $< \rm{T}_{\rm gas} <$ 70 K.

The extended  emission beyond the continuum edge at 200 au is best reproduced by enhanced UV penetration potentially boosted by radial drift, leading to CO and N$_2$ photodesorption as well as increased cosmic ray/X-ray ionization that enhances the abundance of H$_3^+$. Additional processes, including a thermal inversion, variations in the H$_2$ ortho-to-para ratio, and pressure gradients may also contribute. The observed \ntwohplus and \dcoplus morphologies are broadly consistent with estimates of the midplane CO snowline location at 45-85 au, but tighter constraints on the inner disk ionization and temperature structures are needed. 

Overall, this work illustrates that \dcoplus and \ntwohplus emission in protoplanetary disks is shaped by multiple processes, highlighting the need for multi-molecular observations and self-consistent modeling that accounts for these processes and their impact on emission morphologies and disk properties such as snowline locations.

\begin{acknowledgments}
We thank the referee for their helpful comments, which helped improve the manuscript. This research makes use of the following ALMA data: ADS/JAO.ALMA$\#$2016.1.00724.S and ADS/JAO.ALMA$\#$2018.1.01119.S. ALMA is a partnership of ESO (representing its member states), NSF (USA), and NINS (Japan), together with NRC (Canada), MOST and ASIAA (Taiwan), and KASI (Republic of Korea), in cooperation with the Republic of Chile. The Joint ALMA Observatory is operated by ESO, AUI/NRAO, and NAOJ. The National Radio Astronomy Observatory is a facility of the National Science Foundation operated under cooperative agreement by Associated Universities, Inc. A.D. acknowledges support from the NASA FINESST Grant no. 80NSSC24K1470, the John F. Angle Fellowship in Astronomy, and the ACM SIGHPC Computational and Data Science Fellowship.  L.I.C. acknowledges support from the David and Lucille Packard Foundation and NSF grant no. AST2205698. L.I.C. and A.D. acknowledge support from the Research Corporation for Science Advancement Cottrell Fellowship.   This work was also supported by the NKFIH NKKP grant ADVANCED 149943, the NKFIH excellence grant TKP2021-NKTA-64, and the NKFIH \'Elvonal grant KKP 143986. Project no.149943 has been implemented with the support provided by the Ministry of Culture and Innovation of Hungary from the National Research, Development and Innovation Fund, financed under the NKKP ADVANCED funding scheme. AMH is supported by Grant No. ASTR-2307920 from the National Science Foundation.
\end{acknowledgments}
\facility{ALMA}
\software{astropy \citep{astropy}, gofish \citep{GoFish}, vis$\_$sample \citep{loomis2018}, bettermoments \citep{Bettermoments2018}}

\appendix
\vspace{-5mm}
\section{Turbulence}\label{sec:turb}
\begin{figure*}[t!]
\center
\includegraphics[scale=.42]{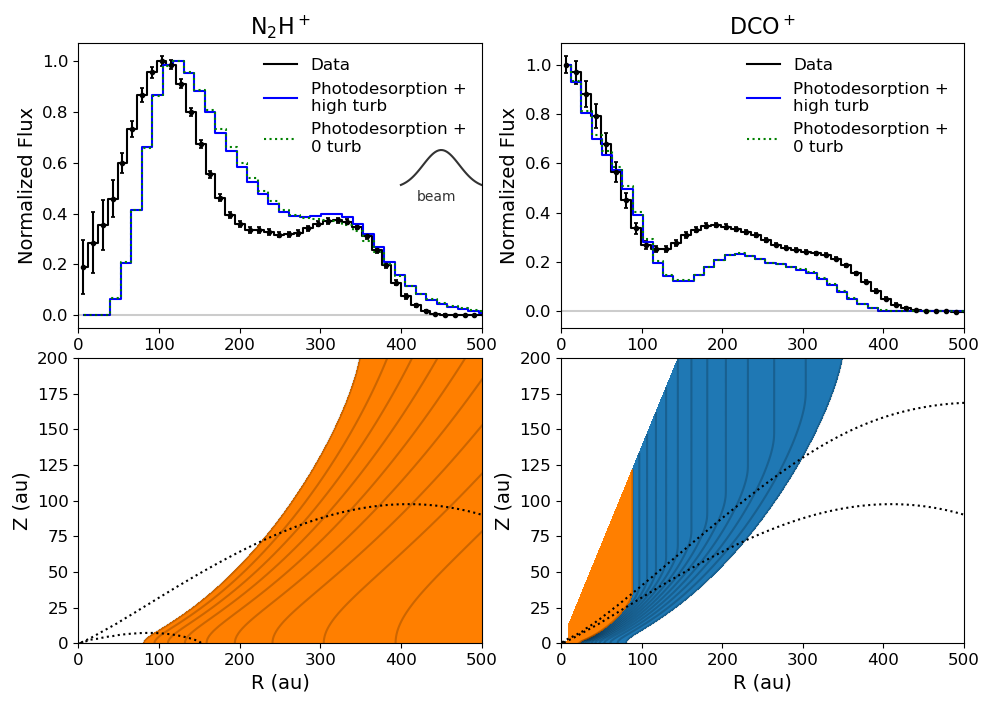}
\includegraphics[scale=.44]{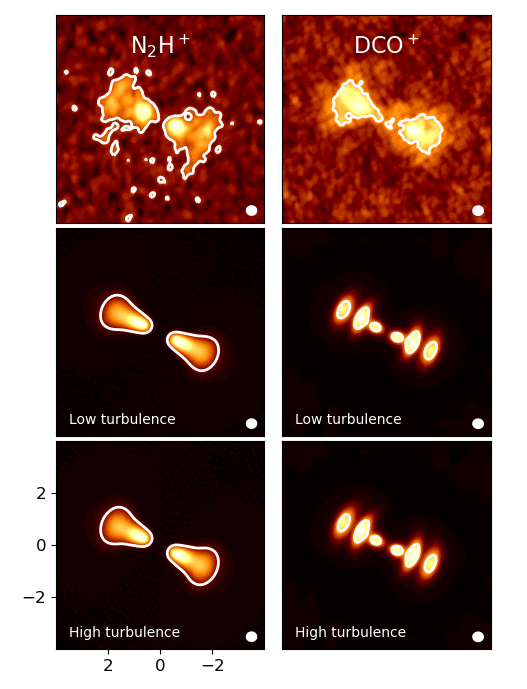}
\includegraphics[scale=.35]{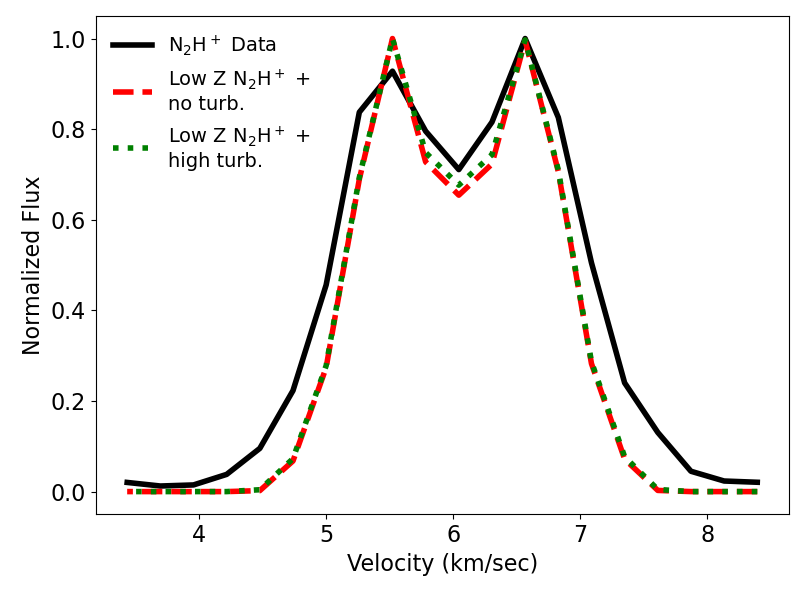}
\includegraphics[scale=.35]{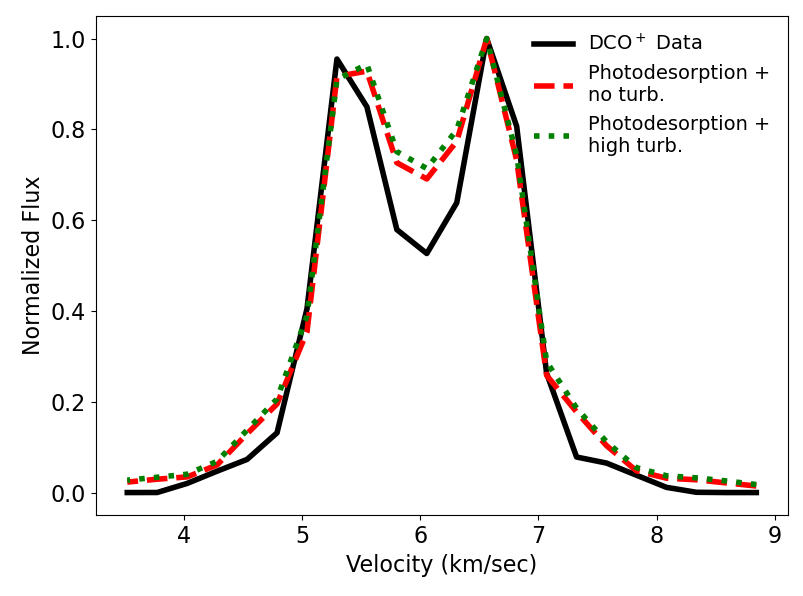}
\caption{Same as row 5 of Figure~\ref{fig:radprofile_all} considering both a model with zero turbulence and a model with turbulence set to 30\% of the local sound speed, comparable to that found from CO emission. The bottom two panels show the spectra for these models. \label{fig:radprofile_turbulence} }
\end{figure*}

We briefly explore the impact of turbulence on our models. We consider the model that provides the best match for \ntwohplus and \dcoplus accounting for photodesorption and ionization, as described in Section~\ref{sec:photod_models}. To this model, we add turbulence equivalent to 30\%\ of the local sound speed. This turbulent velocity is comparable to that found from CO \citep{Flaherty2020} and provides an upper limit on the expected turbulence at the midplane, as models predict that turbulent velocities decrease towards the midplane \citep{Simon2015}. We then run this high-turbulence model through our downhill simplex procedure to determine the abundances and inner radius of the outer ring. Figure~\ref{fig:radprofile_turbulence} shows the resulting radial profiles, channel maps, and spectra. Overall, the inclusion of turbulence does not lead to a significant change in the radial profiles, but there are modest changes in the channel map morphology and the spectra. In the channel maps, there is a slight azimuthal broadening in the outer disk when turbulence is included, and in the spectra the ratio of flux in the line peak to that in the central channel is smaller when turbulence is non-zero. The BIC values (Tables \ref{tab:n2hmodel_params} and \ref{tab:dcomodel_params}) also show a preference for the high turbulence model, as indicated by its lower BIC, suggesting that some level of turbulence near the disk midplane may be present. 

\vspace{10mm}
\section{Model Parameters}
Tables \ref{tab:n2hmodel_params} and \ref{tab:dcomodel_params} show the parameters values for the models shown in this work. The abundances and the inner radii of the  rings were determined using a downhill simplex algorithm. The code is available at \url{https://github.com/kevin-flaherty/DMTau_Chemistry} and version 1.0 of the code used for this paper can be obtained from \url{https://doi.org/10.5281/zenodo.22234211}.

\begin{deluxetable}{cccccccc}[bth!]\label{n2hp_table}
\tablecaption{\ntwohplus Model Parameters\label{tab:n2hmodel_params}}
\tablewidth{0pt}
\tablehead{\colhead{Model} & \colhead{T$_{CO}$ (K)} & \colhead{[$q_{mid}$,$q_{atm}$]}  & \colhead{$\log$(X$_{\text{N}_2\text{H}^+}$)}  &  \colhead{$\log$(ring X$_{\text{N}_2\text{H}^+}$)} & \colhead{R$_{ring}$ (au)} & \colhead{Reduced $\chi^2$} & \colhead{BIC score}}
\decimals
\startdata
CO fiducial & 18 & [-0.371,-0.371] & -11.24 & n/a & n/a & 0.993855 & 15803886\\
            & 20 & [-0.371,-0.371] & -11.24 & n/a & n/a & 0.993792 & 15802879\\
Flat midplane temperature & 18 & [0,-0.371]\tablenotemark{a} & -11.31 & n/a & n/a & 0.993811 & 15803181\\
                          & 20 & [0,-0.371] & -11.31 & n/a & n/a & 0.993805 & 15803086\\
Thermal inversion at 200 au\tablenotemark{b} & 18 & [-0.371,-0.371] & -11.24  & n/a &n/a &  0.993907 & 15804714\\
 & 20 & [-0.371,-0.371] & -11.31 & n/a & n/a &  0.993824 & 158.03388\\
High outer ring\tablenotemark{c} & 18 & [-0.371,-0.371] & -11.24 & -9.90 & 270 & 0.993813 & 15803244\\
                                 & 20 & [-0.371,-0.371] & -11.27 & -9.53 & 270 & 0.993746 & 15802179\\ 
Low outer ring\tablenotemark{d} & 18 & [-0.371,-0.371] & -11.25 & -11.09 & 321 & 0.993788 & 15802853\\ 
                                & 20 & [-0.371,-0.371] & -11.28 & -11.03 & 320 & 0.993715 & 15801686\\
Low outer ring\tablenotemark{d} + turbulence & 18 & [-0.371,-0.371] & -11.28 & -10.99 & 321 & 0.993612 & 15800041 \\
GS25 CO fiducial\tablenotemark{e} & 18 & [-0.5,-0.13] & -11.17 & n/a & n/a & 0.995119 & 15823981\\
                 & 20 & [-0.5,-0.13] & -11.17 & n/a & n/a & 0.994775 & 15818511\\
GS25 + Low outer ring\tablenotemark{d,e} & 18 & [-0.5,-0.13] & -11.21 & -11.52 & 300 & 0.995109 & 15823853\\
                                     & 20 & [-0.5,-0.13] & -11.22 & -11.60 & 312 & 0.994763 & 15818351\\
\enddata
\tablenotetext{a}{At the midplane, the temperature has a flat profile with radius, i.e. $T_{\rm mid} = T_{\rm mid0}$, while the temperature still decreases in the disk atmosphere, i.e., $T_{\rm atm} = T_{\rm atm0} (r/150 au)^{-0.371}$.}
\tablenotetext{b}{A multiplicative factor of 1.3 is applied to the midplane temperature at all radii greater than 200 au.}
\tablenotetext{c}{The high outer ring is placed above the CO photodesorption layer ($0.01 \leq \Sigma_{21} \leq 0.79$).}
\tablenotetext{d}{The low outer ring is placed below the CO photodesorption layer ($3 \leq \Sigma_{21} \leq 100$).}
\tablenotetext{e}{Using the temperature structure from \citet{galloway25}.}
\end{deluxetable}

\begin{deluxetable}{ccccccccc}[htb!]\label{dcop_table}
\tablecaption{\dcoplus Model Parameters\label{tab:dcomodel_params}}
\tablewidth{0pt}
\tablehead{\colhead{Model} & \colhead{T$_{CO}$ (K)} & \colhead{[$q_{mid}$,$q_{atm}$]} & \colhead{\makecell{$\log$\\($\text{X}_{\text{DCO}^+,w})$\tablenotemark{a}} }& \colhead{\makecell{$\log$\\($\text{X}_{\text{DCO}^+,c})$\tablenotemark{a}}}  & \colhead{\makecell{$\log$\\(ring $\text{X}_{\text{DCO}^+}$)}} & \colhead{R$_{ring}$ (au)} & \colhead{Reduced $\chi^2$} & \colhead{BIC score}}
\decimals
\startdata
CO fiducial & 18 & [-0.371,-0.371] & -13.54 & -12.45 & n/a & n/a & 0.929600 & 16454278\\
            & 20 & [-0.371,-0.371] & -14.39 & -11.91 & n/a & n/a & 0.929663 & 16455393\\
Flat midplane temperature & 18 & [0,-0.371]\tablenotemark{b} & -23.58 & -11.96 & n/a & n/a & 0.929194 & 16447092\\
                          & 20 & [0,-0.371] & -24.37 & -11.30 & n/a & n/a & 0.929396 & 16450667\\
Thermal inversion at 200 au\tablenotemark{c} & 18 & [-0.371,-0.371] & -14.01 & -12.31 & n/a & n/a & 0.929420 & 16451092\\
                                             & 20 & [-0.371,-0.371] & -22.48 & -11.85 & n/a & n/a & 0.929622 & 16454668\\
Outer ring\tablenotemark{d} & 18 & [-0.371,-0.371] & -13.27 & -12.60 & -9.50 & 238 & 0.929057 & 16444699\\
                            & 20 & [-0.371,-0.371] & -13.69 & -12.00 & -9.49 & 232 & 0.929077 & 16445053\\
Outer ring\tablenotemark{d}, CP only & 18 & [-0.371,-0.371] & n/a & -12.52 & -9.51 & 238 & 0.929075 & 16445001\\
                                     & 20 & [-0.371,-0.371] & n/a & -11.97 & -9.49 & 233 & 0.929079 & 16445072\\
Outer ring\tablenotemark{d} + turbulence & 18 & [-0.371,-0.371] & -13.31 & -12.62 & -9.49 & 236 & 0.929048 & 16444539\\
GS25 CO fiducial\tablenotemark{e} & 18 & [-0.5,-0.13] & -13.41 & -12.28 & n/a & n/a & 0.929044 & 16444437\\
                 & 20 & [-0.5,-0.13] & -13.40 & -11.91 & n/a & n/a & 0.929247 & 16440862\\
GS25 + Outer ring\tablenotemark{d,e} & 18 & [-0.5,-0.13] & -13.40 & -12.40 & -9.96 & 262 & 0.928915 & 16442154\\
                                   & 20 & [-0.5,-0.13] & -13.41 & -12.08 & -9.83 & 248 & 0.929021 & 16444061\\
\enddata

\tablenotetext{a}{$\text{X}_{\text{DCO}^+,w}$ and $\text{X}_{\text{DCO}^+,c}$ are respectively the abundances of the warm and cold pathway at 50 K and 25 K as defined in Eq. \ref{eq:xdco_warm} and \ref{eq:xdco_cold}.}
\tablenotetext{b}{At the midplane, the temperature has a flat profile with radius, i.e. $T_{\rm mid} = T_{\rm mid0}$, while the temperature still decreases in the disk atmosphere, i.e., $T_{\rm atm} = T_{\rm atm0} (r/150 au)^{-0.371}$.}
\tablenotetext{c}{A multiplicative factor of 1.3 is applied to the midplane temperature at all radii greater than 200 au.}
\tablenotetext{d}{The  outer ring is placed within the CO photodesorption layer ( $0.79 \leq \Sigma_{21} \leq 3$).}
\tablenotetext{e}{Using the temperature structure from \citet{galloway25}.}
\end{deluxetable}

\listofchanges
\end{document}